\documentclass[twoside,twocolumn,english,aps,prl]{revtex4-1}

	\usepackage[usenames,dvipsnames]{xcolor}
	\usepackage{amsmath}
	\usepackage{amsfonts}
	\usepackage{amssymb}
	\usepackage[colorlinks=true,citecolor=blue,linkcolor=red]{hyperref}
	\usepackage{graphicx}
	\usepackage{bbold}					
	\usepackage[makeroom]{cancel}		
	\usepackage{multirow}				
	\usepackage[normalem]{ulem}        
	\usepackage{array}
	\usepackage{pdfpages}
	\makeatletter
	\AtBeginDocument{\let\LS@rot\@undefined}
	\makeatother
	
	\newcolumntype{x}[1]{>{\centering\let\newline\\\arraybackslash\hspace{0pt}}p{#1}}
	
	\renewcommand{\Im}{\operatorname{Im}} 
	
	\DeclareMathAlphabet{\mathbbold}{U}{bbold}{m}{n}

	\DeclareMathOperator{\Tr}{Tr}
	
	\newcounter{subeqn} %
	\makeatletter
	\@addtoreset{subeqn}{equation}
	\makeatother

\definecolor{TB}{rgb}{0,0,0} 

\begin{document}

\title{ Dissipative Floquet Majorana modes in proximity-induced topological superconductors}

\author{Zhesen Yang$^{1}$}
\author{Qinghong Yang$^{2,3}$}
\author{Jiangping Hu$^{1,4,5}$}
\author{Dong E. Liu$^{2,3,6}$}\email{Corresponding to: dongeliu@mail.tsinghua.edu.cn}

\affiliation{$^{1}$Beijing National Laboratory for Condensed Matter Physics, and Institute of Physics, Chinese Academy of Sciences, Beijing 100190, China}

\affiliation{$^{2}$State Key Laboratory of Low Dimensional Quantum Physics, Department of Physics, Tsinghua University, Beijing, 100084, China}

\affiliation{$^{3}$Beijing Academy of Quantum Information Sciences, Beijing 100193, China}

\affiliation{$^{4}$Kavli Institute for Theoretical Sciences, University of Chinese Academy of Sciences, Beijing 100190, China}

\affiliation{$^{5}$South Bay Interdisciplinary Science Center, Dongguan, Guangdong Province, China}

\affiliation{$^{6}$Frontier Science Center for Quantum Information, Beijing 100184, China}

\date{\today}

\begin{abstract}
We study a realistic Floquet topological superconductor, a periodically driven nanowire proximitized to an equilibrium s-wave superconductor. Due to both strong energy and density fluctuations caused from the superconducting proximity effect, the Floquet Majorana wire becomes dissipative. We show that the Floquet band structure is still preserved in this dissipative system. In particular, we find that both the Floquet Majorana zero and $\pi$ modes can no longer be simply described by the Floquet topological band theory. We also propose an effective model to simplify the calculation of the lifetime of these Floquet Majoranas, and find that the lifetime can be engineered by the external driving field. 
\end{abstract}

\maketitle

{\em Introduction.}---Floquet engineering, which controls quantum systems using periodic driving~\cite{RevModPhys.89.011004,doi:10.1146/annurev-conmatphys-031218-013423,2019arXiv190902008R}, is believed to provide a potentially accessible method to realize topological nontrivial band structures and other exotic quantum states~\cite{inoue10,lindner11,kitagawaGF11,Dahlhaus11,jiang11,Kitagawa:2012aa,reynoso12,Liu13,IadecolaPRL13,Fregoso13,Iadecola14,FoaTorres14,Sedrakyan15,KitagawaNC12,RechtsmanNature,Struck13,FloquetClassification,FloquetPeriodicTable,BomantaraPRL18,BomantaraPRB18,PengYangPRB18,BauerPRB19,PhysRevLett.117.087402,PhysRevB.96.195303}. In spite of their success in non-interacting (or isolated) systems, the standard Floquet theorem cannot correctly capture many realistic quantum systems: there is a crossover between a pre-thermal regime~\cite{MoriFloquet,AbaninFloquetPrethermal1,AbaninFloquetPrethermal2} to featureless infinite temperature states~\cite{PONTE2015196} in non-integrable interacting Floquet systems; an open Floquet system usually shows complicated statistical behaviors depending on the details of system-bath couplings~\cite{Hone09,PhysRevB.91.144301,PhysRevX.5.041050,PhysRevB.91.235133}. Certain self-consistent treatments along with realistic conditions are crucial for understanding such elusive non-equilibrium systems.

An example is the Floquet Majorana modes in the periodically driven topological superconductors (SCs)~\cite{jiang11,reynoso12, Liu13, FloquetClassification, BomantaraPRB18}. In most experimental realizable proposals, the topological SC is induced by the proximity effect~\cite{Fu&Kane08,SatoPRL09,Sau10,LutchynPRL10,1DwiresOreg,Sau10}. In the equilibrium case, the low energy physics related to proximity effect can usually be well approximated by an {\em ad hoc} pairing term, e.g., the self-energy correction at zero frequency in the nanowire $\Sigma_{sc}(\omega)\simeq \Sigma_{sc}(\omega=0)=\Delta_{ind}\sigma_y\tau_y$~\cite{PhysRevB.96.014510}; and therefore, the resulting minimal model is equivalent to the intrinsic SC, which has been extensively applied in the literature~\cite{Sau10,LutchynPRL10,1DwiresOreg,Sau10}. However, when the system is under external driving, a realistic framework is to regard the SC as an external bath, which not only provides the tunneling of Cooper pairs but also acts as a dissipative source that renders the periodically driven nanowire reaching a non-equilibrium steady state~\cite{DELPRB17}. More importantly, the zero frequency approximation might no longer be suitable due to the existence of Floquet Majorana $\pi$ modes (FMPMs) at $E=\pm\Omega/2$~\cite{jiang11, Liu13, PhysRevLett.121.076802, PhysRevB.99.094303}, which indicates the physics around $\omega=\pm\Omega/2$ are also important, where $\Omega$ is the driving frequency. This calls for a more realistic minimal model to study the corresponding dissipative Floquet Majoranas and topological phase transitions, and ask whether the Floquet picture is still valid if there exists the interplay between the non-equilibrium conditions and strong dissipations.

\begin{figure}[t]
	\centerline{\includegraphics[height=3.1cm]{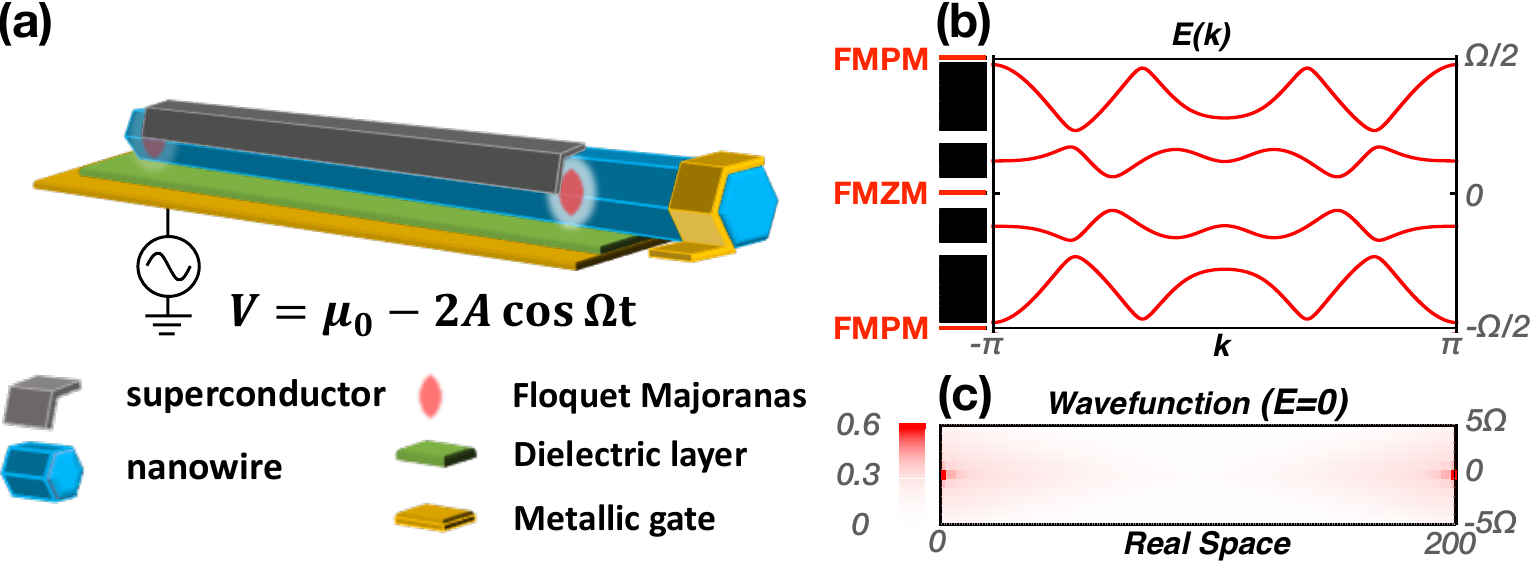}}
	\caption{The closed (or intrinsic) Floquet SC limit. (a) shows the setup. (b) and (c) show the open boundary spectra/Floquet band of Eq.~\ref{E3} under the approximation shown in Eq.~\ref{isc} and the wave-function of the Floquet Majorana zero mode with $E=0$. Note that the left black spectrum in (b) is for an open finite wire with lattice size $N=200$, while the right $k-$dependent spectrum is for a wire with spatially periodic boundary condition. The parameters are chosen $t_0=1,\lambda=1.5,\mu_0=-2,V_z=1.2,A=3/2,\Omega=6,V=0.8,N_F=5$. 
		\label{F1}}
\end{figure}

In this paper, we consider a realistic model based on a periodically driven nanowire proximitized to an equilibrium s-wave superconductor, as shown in Fig.~\ref{F1} (a). Using the Keldysh Green's function formalism~\cite{Wyatt66,Wyatt66,Eliashberg70,Robertson09,Mankowsky14,Galitskii69,Elesin71,Galitskii73,Elesin73,ManiNature02,ZudovPRL03,YangPRL03,AndreevPRL03,Vavilov04,Finkler09,GoldsteinPRB15,TorrePRA13,SiebererPRL13,SiebererPRB14,AltmanPRX15,Sieberer15,MaghrebiPRB16}, we study the physics induced by the self-energy correction beyond the zero frequency approximation; and find that under certain conditions, both Floquet Majorana zero modes (FMZMs) and FMPMs become dissipative and can no longer be predicted by the Floquet topological band theory, in which the pairing term is constant and $\omega$ independent. We also propose a Floquet Majorana poisoning model to simulate and evaluate the lifetime of these dissipative Floquet Majoranas. 

{\em Realistic Floquet proximity SC.}---Consider a periodically driven semiconductor nanowire coupled to a conventional $s$-wave SC as shown in Fig.~\ref{F1} (a) (Refer to~\cite{Mourik2012,Deng2012,Das2012,Churchill2013,Finck2013,Albrecht16,deng2016Majorana,Zhang2017Ballistic,ZhangNN2018} for recent experimental progresses of their static counterpart), the Hamiltonian has three parts and can be written as follows~\cite{jiang11,reynoso12,Liu13}
\begin{equation}
\hat{H}(t)=\hat{H}_{nw}(t)+\hat{H}_{sc}+\hat{H}_c.
\label{E1}\end{equation}
Here $\hat{H}_{nw}(t)=\sum_{k} \hat{\Psi}_k^\dag[(-2t_0\cos k-\mu_0+2A\cos\Omega t)\tau_z+V_z\sigma_z\tau_z+\lambda \sin k\sigma_y\tau_z]\hat{\Psi}_k$ is the Hamiltonian of the nanowire driven by the external lead and $\hat{\Psi}_k=(\hat{c}_{k,\uparrow},\hat{c}_{k,\downarrow},\hat{c}_{-k,\uparrow}^\dag,\hat{c}_{-k,\downarrow}^\dag)^t$, where $\hat{c}_{k,\uparrow/\downarrow}$ annihilate spin-up/down electrons with momentum $k$ in the nanowire, and $\sigma_\mu/\tau_\mu$ represent the Pauli matrices in the spin/Nambu spaces; $\hat{H}_{sc}=\sum_q \hat{\Phi}^\dag_q(\epsilon_q \tau_z-\Delta\sigma_y\tau_y)\hat{\Phi}_q$ is the Hamiltonian of SC bath, and  $\hat{\Phi}_q=(\hat{a}_{q,\uparrow},\hat{a}_{q,\downarrow},\hat{a}_{-q,\uparrow}^\dag,\hat{a}_{-q,\downarrow}^\dag)^t$, where $\hat{a}_{q,\uparrow/\downarrow}$ annihilate spin-up/down electrons with momentum $q$ in the SC bath; $\hat{H}_c=\sum_{k,q,\sigma}(V\hat{c}_{k,\sigma}^\dag\hat{a}_{q,\sigma}+V^*\hat{a}_{q,\sigma}^\dag\hat{c}_{k,\sigma})$ is Hamiltonian describing the nanowire-bath coupling. Here the parameters $\mu_0,V_z,\lambda,\Delta,V$ represent the static chemical potential, Zeeman field, spin-orbit coupling strength, SC order parameter of the SC bath and the nanowire-bath coupling strength, respectively. The external driving is controlled by the amplitude $A$ and frequency $\Omega$. Without loss of generality, $\Delta$ and $V$ are assumed to be real positive numbers.  

It is widely believed that an open Floquet system coupled to an external thermal bath will eventually reach a non-equilibrium steady state, in which the energy absorbed from the external driving field is balanced by the energy flowing out to the environment~\cite{RevModPhys.89.011004,2019arXiv190902008R}. Theoretically, the physical observables in the non-equilibrium steady state can be conveniently dealt with within the framework of Keldysh formalism~\cite{DELPRB17}. To be more precise, spectral properties and distribution functions can be calculated from the retarded and Keldysh components of the Keldysh Green's function. We mainly focus on spectral properties and their quasi-particle lifetimes, so only the retarded component is needed. Using the Floquet theorem, in the supplementary material (SM)~\cite{supp}, we show that when the external SC bath degrees of freedom are integrated out, the retarded component of the Keldysh Green's function has the following form~\cite{note},
\begin{equation}
\underline{G_{nw}^{R}(k,\omega)}=\left[\underline{\omega}-\underline{\mathcal{H}_{eff}(k,\omega)}\right]^{-1},
\label{E2}
\end{equation}
where $\underline{\omega}$ is proportional to the identity matrix and 
\begin{widetext}
	\begin{equation}
	\underline{\mathcal{H}_{eff}(k,\omega)}=\left(\begin{array}{ccccc}
	{\dots} &  &  &  &  \\ 
	& \mathcal{H}_{nw}(k)-\Omega+\Sigma_{sc}(\omega+\Omega) & A\sigma_0\tau_z & 0 &   \\ 
	&  A\sigma_0\tau_z & \mathcal{H}_{nw}(k)+\Sigma_{sc}(\omega) & A\sigma_0\tau_z &   \\ 
	&  0 & A\sigma_0\tau_z & \mathcal{H}_{nw}(k)+\Omega+\Sigma_{sc}(\omega-\Omega) &   \\ 
	&  &  &  &  {\dots} 
	\end{array}\right).
	\label{E3}
	\end{equation}
Here
\begin{equation}
\mathcal{H}_{nw}(k)=(-2t_0\cos k-\mu_0)\tau_z+V_z\sigma_z\tau_z+\lambda \sin k\sigma_y\tau_z,\quad \Sigma_{sc}(\omega)=V^2\left[-(\omega+ i \eta)-\Delta \sigma_{y} \tau_{y}\right]/\sqrt{-(\omega+ i \eta)^{2}+\Delta^{2}},\quad 
\label{E4}
\end{equation}
\end{widetext}
are the static Hamiltonian of the nanowire and the static self-energy correction with $\eta=0^+$~\cite{DELPRB17}, and the object $\underline{X}$ is a matrix with infinite dimension in the Floquet space, whose basis is spanned by the Harmonic functions $e^{-i(\omega+m\Omega)t}$ with $m=0,\pm1,...$~\cite{RevModPhys.89.011004,supp}. In practice, a truncation $N_F$ is necessary for numerics, e.g. $m=0,\pm1,...,\pm N_F$. We emphasize that in the derivation of Eq.~\ref{E2}-\ref{E4}, we have made a physical assumption, i.e. the external SC density of states (DoS) is constant~\cite{supp}. However, the Green's function method we applied is exact without approximation~\cite{PhysRevB.96.014510,Mahan}. In the static limit ($A=0$), the low-energy equilibrium physics of the nanowire can be well described by the following zero frequency approximation
\begin{equation}\begin{aligned}
\mathcal{H}_{A=0}(k)\simeq\mathcal{H}_{nw}(k)+\Sigma_{sc}(\omega=0)= \mathcal{H}_{nw}(k)+\Delta_{ind}\sigma_y\tau_y,
\end{aligned}
\label{static}\end{equation}
where $\Delta_{ind}=-V^2$. Physically, the Hamiltonian Eq.~\ref{static} is equivalent to the intrinsic SC with order parameter $\Delta_{ind}$~\cite{PhysRevB.96.014510}. When $V_z^2>(\pm2t_0+\mu)^2+\Delta_{ind}^2$, the system can have Majorana zero modes at two boundaries~\cite{LutchynPRL10,1DwiresOreg}. Under external drivings, when we take the following intrinsic SC approximation 
\begin{equation}
\Sigma_{sc}(\omega+m\Omega)\simeq \Delta_{ind}\sigma_y\tau_y, \quad m=-N_F,...,N_F,
\label{isc}
\end{equation}
Eq.~\ref{E3} reduces to the Floquet Hamiltonian studied in Ref.~\cite{jiang11,Liu13}. The emergence or absence of Floquet Majoranas can be well described by the Floquet topological band theory~\cite{inoue10,lindner11,kitagawaGF11,Dahlhaus11,jiang11,PhysRevB.96.195303}. Fig.~\ref{F1} (b) shows an example of the Floquet band structure and open boundary spectra with $N_F=5$ (other parameters are shown in the caption of Fig.~\ref{F1}). Both FMZMs and FMPMs do exist in the open boundary spectrum~\footnote{We note that in the static case with the parameters we chosen, the system is topological nontrivial and has Majorana zero modes on the boundary.}. The corresponding wave-function of the FMZM with $E=0$ is plotted in Fig.~\ref{F1} (c). One can also notice that the FMZM is localized not only in the real space but also in the Floquet space. Due to the translational symmetry of the Floquet Hamiltonian, the FMZMs with $E=n\Omega$ must be localized at Floquet sites $n\Omega$~\cite{PhysRevB.98.220509}.  

Now we explain why the above intrinsic SC approximation Eq.~\ref{isc} can no longer be applied in Floquet proximity topological SCs. We first discuss the mathematical meaning of Eq.~\ref{isc}. From Eq.~\ref{E4},  $\Sigma_{sc}(\omega+m\Omega)=-V^2(\sigma_{y} \tau_{y}+\delta)/\sqrt{1-\delta^{2}}$, where $\delta=(\omega+m\Omega+i\eta)/\Delta$, one can notice that only when $\delta\rightarrow0$, the approximation Eq.~\ref{E4} can be applied. This requires $2\Delta\gg (2N_F+1)\Omega$ for a fixed $N_F$. Obviously, this requirement cannot be achieved in experiments. Therefore, the self-energy can no longer be approximated by a constant and $\omega$ independent term. Indeed, from Eq.~\ref{E4}, when $|\omega|>\Delta$, $\Sigma_{sc}(\omega)=iV^2[-|\omega| -\Delta\sigma_y\tau_y]/\sqrt{\omega^2-\Delta^2}$ becomes pure imaginary, which plays the role of effective dissipations in the nanowire. Therefore, in order to investigate the lifetime of the corresponding Floquet Majoranas, a suitable treatment of the self-energy correction is necessary. 

\begin{figure}[t]
	\centerline{\includegraphics[height=7cm]{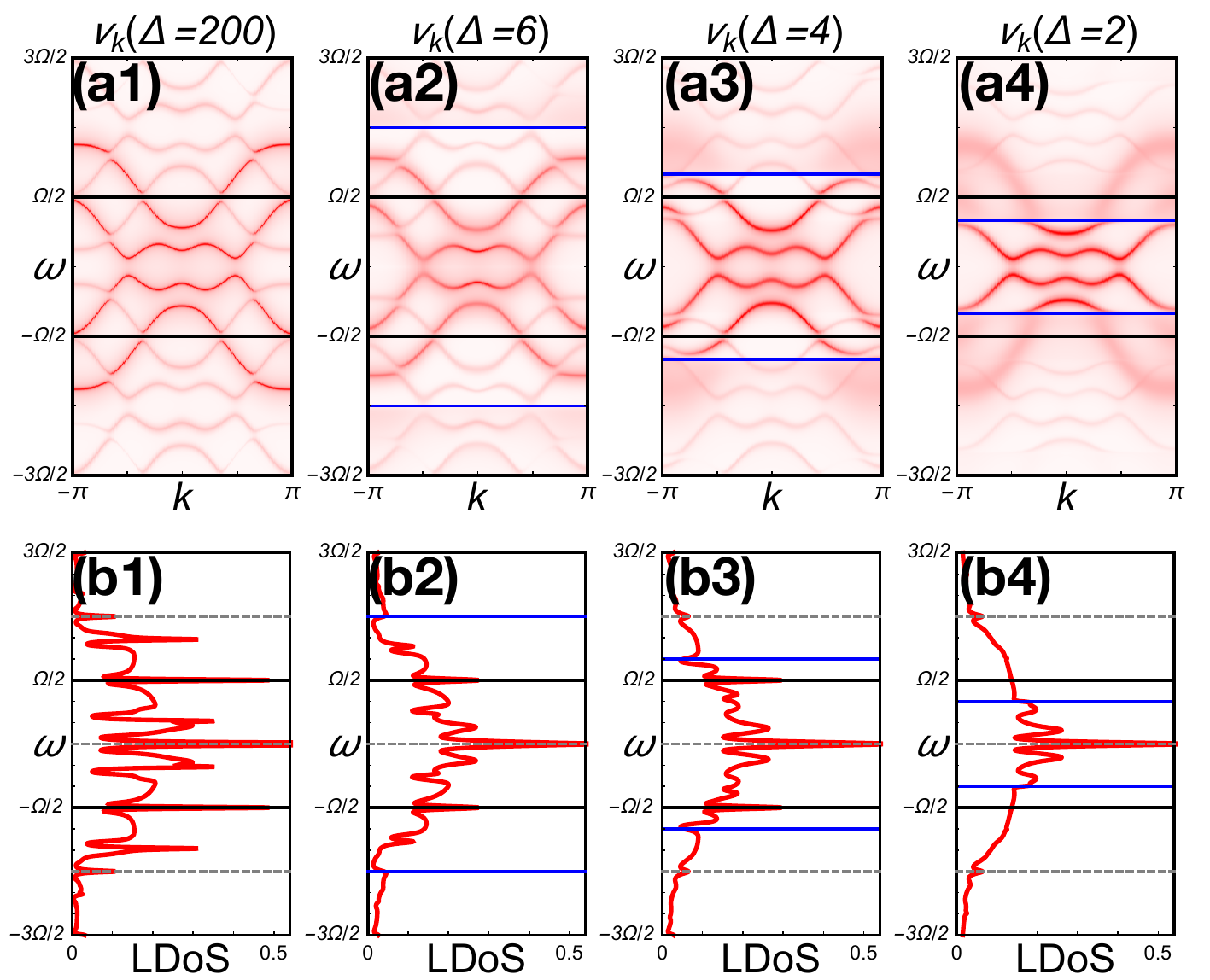}}
	\caption{$\nu_k(\omega)$ (periodic boundary condition) and LDoS (open boundary condition) at the ends of the nanowire. The parameters are chosen the same as Fig.~\ref{F1} except $\eta=0.05$ and different values of $\Delta$ (blue lines) shown above. (a1) shows the same band structure as Fig.~\ref{F1} (c). When $\Delta$ decreases, the bands in (a2)-(a4) change dramatically due to the existence of  non-linear $\omega$ terms in the self-energy correction. When $2\Delta<\Omega$, the FMPMs are destroyed. 
		\label{F2}}
\end{figure}

{\em Periodic boundary condition.}---In order to investigate the role of the  $\omega$-dependent self-energy correction $\Sigma_{sc}(\omega)$, we first apply the numerical calculation of the time-averaged momentum resolved DoS $\nu_k(\omega)$ with spatially periodic boundary condition as shown in Fig.~\ref{F2} (a), where 
\begin{equation}\begin{aligned}
&\nu_k(\omega)=-\frac{1}{\pi}\Tr\Im\left[\underline{G_{nw}^{R}(k,\omega)}\right]_{00},
\label{E6}
\end{aligned}\end{equation}
and the subscript $00$ represents the $00$-Floquet-entry. For example, in Eq.~\ref{E3}, $\left[\underline{\mathcal{H}_{eff}(k,\omega)}\right]_{00}=\mathcal{H}_{nw}(k)+\Sigma_{sc}(\omega)$. As discussed above, when $\Delta\gg(2N_F+1)\Omega$,  $\nu_k(\omega)$ can be approximately described by the Floquet band theory in the intrinsic SC limit with $\Delta_{ind}=-V^2$, whose band structure has been shown in Fig.~\ref{F1} (b). Comparing Fig.~\ref{F1} (b) with Fig.~\ref{F2} (a1), one can find that the approximation Eq.~\ref{isc} works well when $\Delta=200$~\footnote{We note that in Fig.~\ref{F2} (a), the parameters $\Delta=200$, $N_F=5$, $\Omega=6$ satisfy the condition $\Delta\gg(2N_F+1)\Omega$.}. With the decreasing of $\Delta$ (blue lines), sharp features in the spectrum continuously broaden due to the dissipation effect. More interestingly, when $2\Delta$ is smaller than $\Omega$, as shown in Fig.~\ref{F2} (a4), $\nu_k(\omega)$ even exhibits discontinue behavior at $\omega=\pm\Delta$ due to the singularity of $\Sigma_{sc}(\omega=\Delta)$. This is a strong nonlinear self-energy effect, which will kill FMPMs as shown later. 

{\em Open boundary condition.}---We now turn to the discussion of Floquet Majoranas. In order to characterize them, we numerically calculate the time-averaged local DoS (LDoS) at the end of the nanowire with open boundary conditions, based on the recursive Green's function method~\cite{Thouless_1981,PhysRevLett.47.882,DROUVELIS2006741} (refer to SM~\cite{supp} for Floquet systems). As shown in Fig.~\ref{F2} (b1)-(b4), numerical results verify our observations obtained from the band behavior in (a1)-(a4), namely, the killing of FMPMs at $\omega=(2m+1)\Omega/2$. 
To be more precise, when $\Delta$ decreases to $2\Delta<\Omega$, FMPMs are destroyed as shown in (b1)-(b4). This can be understood from the gap closing at $\omega=\pm\Omega/2$ shown in (a4) due to the nonlinear self-energy corrections. Therefore, the FMPM can no longer be described by the Floquet topological band theory under the intrinsic SC limit shown in Fig.~\ref{F1} (b) and Eq.~\ref{isc}. Although in our example, FMZMs are not sensitive to the deceasing of $\Delta$, in the SM, we will show that the topological phase transition of FMZMs is also beyond the description of Floquet topological band theory. We finally note that in the numerical results in (b1)-(b4), for those FMZMs outside the SC gap, their peak heights are very tiny comparing with that inside the SC gap. So, it is important to check whether the lifetime of these Floquet Majoranas in different Floquet Brillouin zones (FBZs)~\footnote{Since the Floquet systems only have discrete time translational symmetry, the energy is only conserved up to $m\Omega$. Therefore, similar to the crystal momentum, one can define the Floquet Brillouin zones for the quasi-energies as follows $\omega\in[(m-1/2)\Omega,(m+1/2)\Omega]$ for $m=-N_F,...,N_F$. When $m=0$, $\omega\in[-\Omega/2,\Omega/2]$ defines the so-called first Floquet Brillouin zones.} are identical or distinct. If their lifetimes are identical, the dissipation-modified Floquet bands are still valid  along with a constant finite lifetime acquired from dissipations. 

\begin{figure}[t]
	\centerline{\includegraphics[height=8.5cm]{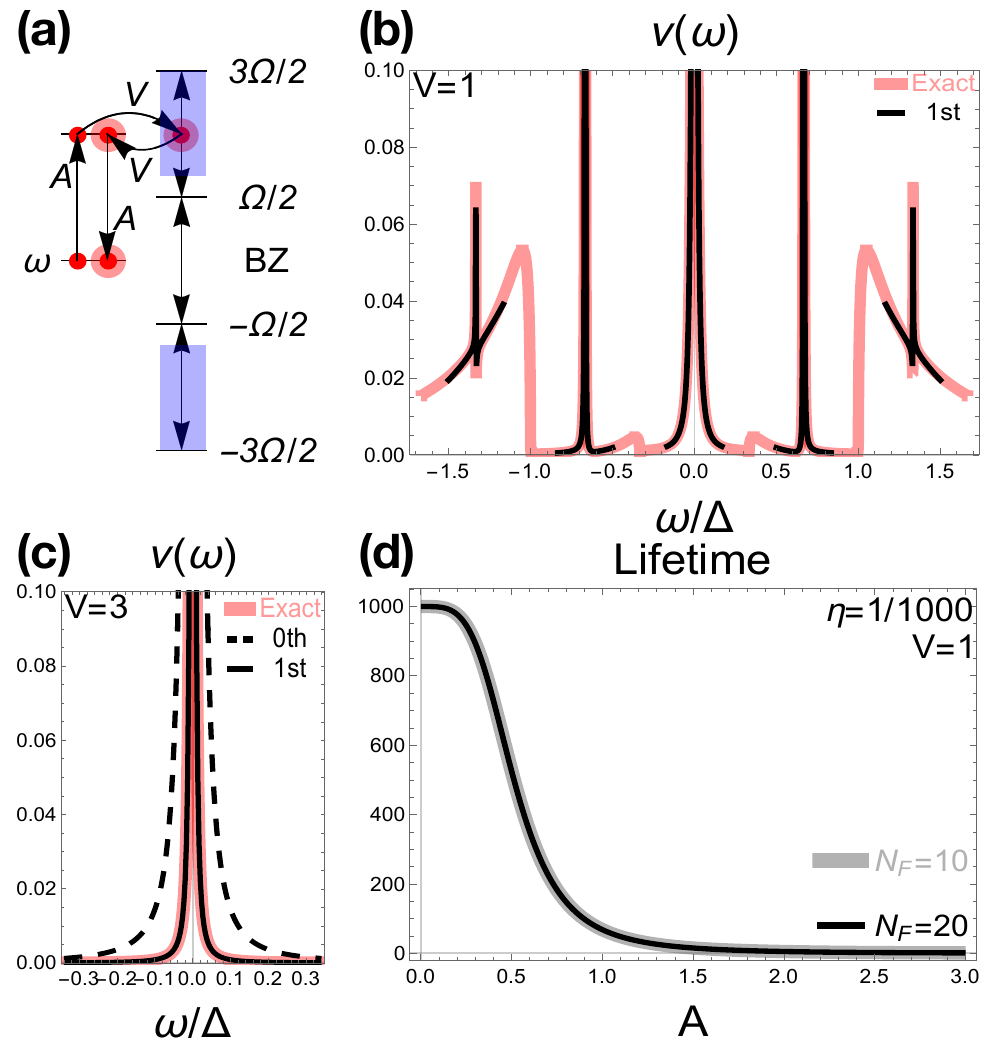}}
	\caption{Floquet Majorana poisoning model. (a) shows the first order procedure. (b) shows the comparison of the exact result and first order approximation of $\nu(\omega)$ in Eq.~\ref{E9} with $\Omega=2$, $\Delta=3$, $A=1/2$, $\eta=1/1000$. (c) show the exact, zeroth and first order numerical calculations. (d) shows the relation between the lifetime and $A$ with the same parameters shown above. 
		\label{F3}}
\end{figure}

{\em Floquet Majorana poisoning model.}---In order to illustrate how the Floquet picture is modified by dissipations provided by the SC, we propose a Floquet Majorana poisoning model, as shown in Fig.~\ref{F3} (a). We emphasize that the Floquet Majorana poisoning model can also be applied to estimate the lifetime of Floquet Majoranas in the above realistic model~\cite{supp}. This model describes a boundary isolated Majorana coupled to a dissipative gapped bath and driven by an external field with frequency $\Omega$ and amplitude $A$. The corresponding $\omega$-dependent Hamiltonian in the Floquet space is
\begin{equation}
	\underline{H(\omega)}=\left(\begin{array}{ccccc}
		{..} &  &  &  &  \\ 
		& \Sigma(\omega+\Omega)-\Omega & A & 0 &   \\ 
		&  A & \Sigma(\omega) & A &   \\ 
		&  0 & A & \Sigma(\omega-\Omega)+\Omega &   \\ 
		&  &  &  &  {..} 
	\end{array}\right),
	\label{E8}
\end{equation}
where $\Sigma(\omega)=-V^2(\omega+ i \eta)/\sqrt{-(\omega+ i \eta)^{2}+\Delta^{2}}$ with $\eta=0^+$, and $V$ is the Majorana-SC bath coupling strength. The time-averaged DoS $\nu(\omega)$ can be calculated from the $00$-Floquet-entry of the retarded Green's function $\underline{G^R(\omega)}=[\underline{\omega}-\underline{H(\omega)}]^{-1}$, i.e.,
\begin{equation}
\nu(\omega)=-\frac{1}{\pi}\Im\left[\underline{G^R(\omega)}\right]_{00}=-\frac{1}{\pi}\Im\left[\frac{1}{\underline{\omega}-\underline{H(\omega)}}\right]_{00}.
\label{E9}
\end{equation}
As shown in Fig.~\ref{F3} (a), although the zero energy Majorana mode is not directly coupled to the bath due to the SC gap, the mode can be excited (or de-excited) to higher FBZs (which directly couples to the SC bath), and cause finite dissipation or poisoning. 

As shown in Fig.~\ref{F3} (b) and (c), the exact results of $\nu(\omega)$ are plotted with the red lines. One can find at each FBZ center $\omega=m\Omega$, there exists a Floquet Majorana peak. However, in the SM, we show that there only exists a single pole of $G^R(\omega)$, which is at $\omega=i\eta$. This means the lifetimes of the Floquet Majoranas at $\omega=\pm m\Omega$ cannot be defined using the traditional method, i.e., finding the imaginary part of the pole of $G^R(\omega)$. However, we find that these Floquet Majorana peaks at $\omega=m\Omega$ can be well fitted by the first order $\omega$-expansion of $\underline{\omega}-\underline{H(\omega)}$ around $\omega=m\Omega$. As shown in Fig.~\ref{F3} (b)-(c), the first order results are shown with solid black lines. As a comparison, we also plot the zeroth order result, i.e., $1/(\underline{\omega}-\underline{H(m\Omega)})$, in (c) with dashed black lines. One can notice that the zeroth order breaks down, while the first order approximation works well. 
We also develop a method to compute the quasi-particle lifetime in the open Floquet quantum systems (refer to SM~\cite{supp} for more details). Here, we only briefly list the main steps: starting from the expansion of the inverse of the Floquet Green's function around $\omega_0$,
\begin{equation}
\det\left[\underline{G^{R,-1}(\omega)}\right]\simeq f_0(\omega_0)+f_1(\omega_0)\delta\omega+o(\delta\omega^2),
\end{equation}
where $\delta\omega=\omega-\omega_0$, and the superscript $-1$ represents the inverse of the retarded Green's function, the lifetime of the quasiparticle around $\omega=\omega_0$ is 
\begin{equation}
\tau_{\omega_0}=\Gamma_{\omega_0}^{-1}=-\frac{1}{\Im[f_0/f_1]},
\end{equation}
and a numerical plot is shown in Fig.~\ref{F3} (d). In addition, due to the discrete time translational symmetry, we have $\det\left[\underline{G^{R,-1}(\omega)}\right]=\det\left[\underline{G^{R,-1}(\omega+m\Omega)}\right]$. This means the expansion of $\det\left[\underline{G^{R,-1}(\omega)}\right]$ around $\omega=m\Omega$ must share the same expression coefficients. Therefore, their lifetimes must be the same, i.e. $\tau|_{\omega=0}=\tau|_{\omega=m\Omega}$. This is consistent with the intuition since the total Hamiltonian does not break the discrete time translational symmetry. In addition, the differences between Floquet Majorana DoS shown in different FBZs (see Fig.~\ref{F2} (b)) come from the Floquet space localization nature of the wavefunction instead of their lifetime difference (refer to the last part of SM~\cite{supp} for more details). In the SM, we also provide a comparison of the FMZM peaks between the Floquet Majorana poisoning model and realistic model.

{\em Discussion and conclusion.}---Our work reveals the crucial role of self-energy correction in the Floquet proximity SC. It is widely believed that the self-energy correction at zero frequency can be described by an effective non-Hermitian Hamiltonian\cite{2017arXiv170805841K,PhysRevB.99.201107,PhysRevLett.121.026403,PhysRevB.98.035141,PhysRevB.99.041116,2020arXiv200302219Y}. The corresponding real and imaginary parts of the eigenvalues can be regarded as the renormalized band structure and quasiparticle lifetime. However, in our model, the linear $\omega$-term of the self-energy is important for the definition of lifetime in the Floquet proximity SC. Further numerical results suggest that the Floquet Majorana lifetime can be tuned by the external field as shown in Fig.~\ref{F3} (d).

In summary, we have shown that in reality Floquet proximity SCs, the intrinsic SC approximation (Eq.~\ref{isc}) is not suitable in understanding the lifetimes of Floquet Majoranas and the their topological features. The difficulty comes from the linear and non-linear effects of the self-energy correction, which can poison and kill Floquet Majoranas, respectively. However, dissipation structures in different FBZs are identical; and therefore, the Floquet picture is still valid in the presence of dissipations.

\begin{acknowledgments} 
D.E.L thanks Roman Lutchyn and Alex Levchenko for the inspired discussions to form the initial motivation of the project. The work is supported by National Science Foundation of China (Grant No. NSFC-11974198, Grant No. NSFC11888101),  the Ministry of Science and Technology of China 973 program
(No. 2017YFA0303100), the Strategic Priority Research Program of CAS (Grant
No.XDB28000000). D.E.L also acknowledge the support from Beijing Academy of Quantum Information Sciences.

\end{acknowledgments}

\bibliography{FloquetK1,FloquetETH_OPEN,DissipativeMaj,FloquetMajoranas,aGBZ}

\begin{thebibliography}{95}%
\makeatletter
\providecommand \@ifxundefined [1]{%
 \@ifx{#1\undefined}
}%
\providecommand \@ifnum [1]{%
 \ifnum #1\expandafter \@firstoftwo
 \else \expandafter \@secondoftwo
 \fi
}%
\providecommand \@ifx [1]{%
 \ifx #1\expandafter \@firstoftwo
 \else \expandafter \@secondoftwo
 \fi
}%
\providecommand \natexlab [1]{#1}%
\providecommand \enquote  [1]{``#1''}%
\providecommand \bibnamefont  [1]{#1}%
\providecommand \bibfnamefont [1]{#1}%
\providecommand \citenamefont [1]{#1}%
\providecommand \href@noop [0]{\@secondoftwo}%
\providecommand \href [0]{\begingroup \@sanitize@url \@href}%
\providecommand \@href[1]{\@@startlink{#1}\@@href}%
\providecommand \@@href[1]{\endgroup#1\@@endlink}%
\providecommand \@sanitize@url [0]{\catcode `\\12\catcode `\$12\catcode
  `\&12\catcode `\#12\catcode `\^12\catcode `\_12\catcode `\%12\relax}%
\providecommand \@@startlink[1]{}%
\providecommand \@@endlink[0]{}%
\providecommand \url  [0]{\begingroup\@sanitize@url \@url }%
\providecommand \@url [1]{\endgroup\@href {#1}{\urlprefix }}%
\providecommand \urlprefix  [0]{URL }%
\providecommand \Eprint [0]{\href }%
\providecommand \doibase [0]{http://dx.doi.org/}%
\providecommand \selectlanguage [0]{\@gobble}%
\providecommand \bibinfo  [0]{\@secondoftwo}%
\providecommand \bibfield  [0]{\@secondoftwo}%
\providecommand \translation [1]{[#1]}%
\providecommand \BibitemOpen [0]{}%
\providecommand \bibitemStop [0]{}%
\providecommand \bibitemNoStop [0]{.\EOS\space}%
\providecommand \EOS [0]{\spacefactor3000\relax}%
\providecommand \BibitemShut  [1]{\csname bibitem#1\endcsname}%
\let\auto@bib@innerbib\@empty
\bibitem [{\citenamefont {Eckardt}(2017)}]{RevModPhys.89.011004}%
  \BibitemOpen
  \bibfield  {author} {\bibinfo {author} {\bibfnamefont {A.}~\bibnamefont
  {Eckardt}},\ }\href {\doibase 10.1103/RevModPhys.89.011004} {\bibfield
  {journal} {\bibinfo  {journal} {Rev. Mod. Phys.}\ }\textbf {\bibinfo {volume}
  {89}},\ \bibinfo {pages} {011004} (\bibinfo {year} {2017})}\BibitemShut
  {NoStop}%
\bibitem [{\citenamefont {Oka}\ and\ \citenamefont
  {Kitamura}(2019)}]{doi:10.1146/annurev-conmatphys-031218-013423}%
  \BibitemOpen
  \bibfield  {author} {\bibinfo {author} {\bibfnamefont {T.}~\bibnamefont
  {Oka}}\ and\ \bibinfo {author} {\bibfnamefont {S.}~\bibnamefont {Kitamura}},\
  }\href {\doibase 10.1146/annurev-conmatphys-031218-013423} {\bibfield
  {journal} {\bibinfo  {journal} {Annual Review of Condensed Matter Physics}\
  }\textbf {\bibinfo {volume} {10}},\ \bibinfo {pages} {387} (\bibinfo {year}
  {2019})},\ \Eprint
  {http://arxiv.org/abs/https://doi.org/10.1146/annurev-conmatphys-031218-013423}
  {https://doi.org/10.1146/annurev-conmatphys-031218-013423} \BibitemShut
  {NoStop}%
\bibitem [{\citenamefont {{Rudner}}\ and\ \citenamefont
  {{Lindner}}(2019)}]{2019arXiv190902008R}%
  \BibitemOpen
  \bibfield  {author} {\bibinfo {author} {\bibfnamefont {M.~S.}\ \bibnamefont
  {{Rudner}}}\ and\ \bibinfo {author} {\bibfnamefont {N.~H.}\ \bibnamefont
  {{Lindner}}},\ }\href@noop {} {\bibfield  {journal} {\bibinfo  {journal}
  {arXiv e-prints}\ ,\ \bibinfo {eid} {arXiv:1909.02008}} (\bibinfo {year}
  {2019})},\ \Eprint {http://arxiv.org/abs/1909.02008} {arXiv:1909.02008
  [cond-mat.mes-hall]} \BibitemShut {NoStop}%
\bibitem [{\citenamefont {Inoue}\ and\ \citenamefont {Tanaka}(2010)}]{inoue10}%
  \BibitemOpen
  \bibfield  {author} {\bibinfo {author} {\bibfnamefont {J.-i.}\ \bibnamefont
  {Inoue}}\ and\ \bibinfo {author} {\bibfnamefont {A.}~\bibnamefont {Tanaka}},\
  }\href {\doibase 10.1103/PhysRevLett.105.017401} {\bibfield  {journal}
  {\bibinfo  {journal} {Phys. Rev. Lett.}\ }\textbf {\bibinfo {volume} {105}},\
  \bibinfo {pages} {017401} (\bibinfo {year} {2010})}\BibitemShut {NoStop}%
\bibitem [{\citenamefont {Lindner}\ \emph {et~al.}(2011)\citenamefont
  {Lindner}, \citenamefont {Refael},\ and\ \citenamefont
  {Galitski}}]{lindner11}%
  \BibitemOpen
  \bibfield  {author} {\bibinfo {author} {\bibfnamefont {N.~H.}\ \bibnamefont
  {Lindner}}, \bibinfo {author} {\bibfnamefont {G.}~\bibnamefont {Refael}}, \
  and\ \bibinfo {author} {\bibfnamefont {V.}~\bibnamefont {Galitski}},\ }\href
  {\doibase doi:10.1038/nphys1926} {\bibfield  {journal} {\bibinfo  {journal}
  {Nature Physics}\ }\textbf {\bibinfo {volume} {7}},\ \bibinfo {pages} {490}
  (\bibinfo {year} {2011})}\BibitemShut {NoStop}%
\bibitem [{\citenamefont {Kitagawa}\ \emph {et~al.}(2011)\citenamefont
  {Kitagawa}, \citenamefont {Oka}, \citenamefont {Brataas}, \citenamefont
  {Fu},\ and\ \citenamefont {Demler}}]{kitagawaGF11}%
  \BibitemOpen
  \bibfield  {author} {\bibinfo {author} {\bibfnamefont {T.}~\bibnamefont
  {Kitagawa}}, \bibinfo {author} {\bibfnamefont {T.}~\bibnamefont {Oka}},
  \bibinfo {author} {\bibfnamefont {A.}~\bibnamefont {Brataas}}, \bibinfo
  {author} {\bibfnamefont {L.}~\bibnamefont {Fu}}, \ and\ \bibinfo {author}
  {\bibfnamefont {E.}~\bibnamefont {Demler}},\ }\href {\doibase
  10.1103/PhysRevB.84.235108} {\bibfield  {journal} {\bibinfo  {journal} {Phys.
  Rev. B}\ }\textbf {\bibinfo {volume} {84}},\ \bibinfo {pages} {235108}
  (\bibinfo {year} {2011})}\BibitemShut {NoStop}%
\bibitem [{\citenamefont {Dahlhaus}\ \emph {et~al.}(2011)\citenamefont
  {Dahlhaus}, \citenamefont {Edge}, \citenamefont {Tworzyd\l{}o},\ and\
  \citenamefont {Beenakker}}]{Dahlhaus11}%
  \BibitemOpen
  \bibfield  {author} {\bibinfo {author} {\bibfnamefont {J.~P.}\ \bibnamefont
  {Dahlhaus}}, \bibinfo {author} {\bibfnamefont {J.~M.}\ \bibnamefont {Edge}},
  \bibinfo {author} {\bibfnamefont {J.}~\bibnamefont {Tworzyd\l{}o}}, \ and\
  \bibinfo {author} {\bibfnamefont {C.~W.~J.}\ \bibnamefont {Beenakker}},\
  }\href {\doibase 10.1103/PhysRevB.84.115133} {\bibfield  {journal} {\bibinfo
  {journal} {Phys. Rev. B}\ }\textbf {\bibinfo {volume} {84}},\ \bibinfo
  {pages} {115133} (\bibinfo {year} {2011})}\BibitemShut {NoStop}%
\bibitem [{\citenamefont {Jiang}\ \emph {et~al.}(2011)\citenamefont {Jiang},
  \citenamefont {Kitagawa}, \citenamefont {Alicea}, \citenamefont {Akhmerov},
  \citenamefont {Pekker}, \citenamefont {Refael}, \citenamefont {Cirac},
  \citenamefont {Demler}, \citenamefont {Lukin},\ and\ \citenamefont
  {Zoller}}]{jiang11}%
  \BibitemOpen
  \bibfield  {author} {\bibinfo {author} {\bibfnamefont {L.}~\bibnamefont
  {Jiang}}, \bibinfo {author} {\bibfnamefont {T.}~\bibnamefont {Kitagawa}},
  \bibinfo {author} {\bibfnamefont {J.}~\bibnamefont {Alicea}}, \bibinfo
  {author} {\bibfnamefont {A.~R.}\ \bibnamefont {Akhmerov}}, \bibinfo {author}
  {\bibfnamefont {D.}~\bibnamefont {Pekker}}, \bibinfo {author} {\bibfnamefont
  {G.}~\bibnamefont {Refael}}, \bibinfo {author} {\bibfnamefont {J.~I.}\
  \bibnamefont {Cirac}}, \bibinfo {author} {\bibfnamefont {E.}~\bibnamefont
  {Demler}}, \bibinfo {author} {\bibfnamefont {M.~D.}\ \bibnamefont {Lukin}}, \
  and\ \bibinfo {author} {\bibfnamefont {P.}~\bibnamefont {Zoller}},\ }\href
  {\doibase 10.1103/PhysRevLett.106.220402} {\bibfield  {journal} {\bibinfo
  {journal} {Phys. Rev. Lett.}\ }\textbf {\bibinfo {volume} {106}},\ \bibinfo
  {pages} {220402} (\bibinfo {year} {2011})}\BibitemShut {NoStop}%
\bibitem [{\citenamefont {Kitagawa}\ \emph
  {et~al.}(2012{\natexlab{a}})\citenamefont {Kitagawa}, \citenamefont {Broome},
  \citenamefont {Fedrizzi}, \citenamefont {Rudner}, \citenamefont {Berg},
  \citenamefont {Kassal}, \citenamefont {Aspuru-Guzik}, \citenamefont
  {Demler},\ and\ \citenamefont {White}}]{Kitagawa:2012aa}%
  \BibitemOpen
  \bibfield  {author} {\bibinfo {author} {\bibfnamefont {T.}~\bibnamefont
  {Kitagawa}}, \bibinfo {author} {\bibfnamefont {M.~A.}\ \bibnamefont
  {Broome}}, \bibinfo {author} {\bibfnamefont {A.}~\bibnamefont {Fedrizzi}},
  \bibinfo {author} {\bibfnamefont {M.~S.}\ \bibnamefont {Rudner}}, \bibinfo
  {author} {\bibfnamefont {E.}~\bibnamefont {Berg}}, \bibinfo {author}
  {\bibfnamefont {I.}~\bibnamefont {Kassal}}, \bibinfo {author} {\bibfnamefont
  {A.}~\bibnamefont {Aspuru-Guzik}}, \bibinfo {author} {\bibfnamefont
  {E.}~\bibnamefont {Demler}}, \ and\ \bibinfo {author} {\bibfnamefont {A.~G.}\
  \bibnamefont {White}},\ }\href {\doibase 10.1038/ncomms1872} {\bibfield
  {journal} {\bibinfo  {journal} {Nature Communications}\ }\textbf {\bibinfo
  {volume} {3}},\ \bibinfo {pages} {882} (\bibinfo {year}
  {2012}{\natexlab{a}})}\BibitemShut {NoStop}%
\bibitem [{\citenamefont {Reynoso}\ and\ \citenamefont
  {Frustaglia}(2013)}]{reynoso12}%
  \BibitemOpen
  \bibfield  {author} {\bibinfo {author} {\bibfnamefont {A.~A.}\ \bibnamefont
  {Reynoso}}\ and\ \bibinfo {author} {\bibfnamefont {D.}~\bibnamefont
  {Frustaglia}},\ }\href {\doibase 10.1103/PhysRevB.87.115420} {\bibfield
  {journal} {\bibinfo  {journal} {Phys. Rev. B}\ }\textbf {\bibinfo {volume}
  {87}},\ \bibinfo {pages} {115420} (\bibinfo {year} {2013})}\BibitemShut
  {NoStop}%
\bibitem [{\citenamefont {Liu}\ \emph {et~al.}(2013)\citenamefont {Liu},
  \citenamefont {Levchenko},\ and\ \citenamefont {Baranger}}]{Liu13}%
  \BibitemOpen
  \bibfield  {author} {\bibinfo {author} {\bibfnamefont {D.~E.}\ \bibnamefont
  {Liu}}, \bibinfo {author} {\bibfnamefont {A.}~\bibnamefont {Levchenko}}, \
  and\ \bibinfo {author} {\bibfnamefont {H.~U.}\ \bibnamefont {Baranger}},\
  }\href {\doibase 10.1103/PhysRevLett.111.047002} {\bibfield  {journal}
  {\bibinfo  {journal} {Phys. Rev. Lett.}\ }\textbf {\bibinfo {volume} {111}},\
  \bibinfo {pages} {047002} (\bibinfo {year} {2013})}\BibitemShut {NoStop}%
\bibitem [{\citenamefont {Iadecola}\ \emph {et~al.}(2013)\citenamefont
  {Iadecola}, \citenamefont {Campbell}, \citenamefont {Chamon}, \citenamefont
  {Hou}, \citenamefont {Jackiw}, \citenamefont {Pi},\ and\ \citenamefont
  {Kusminskiy}}]{IadecolaPRL13}%
  \BibitemOpen
  \bibfield  {author} {\bibinfo {author} {\bibfnamefont {T.}~\bibnamefont
  {Iadecola}}, \bibinfo {author} {\bibfnamefont {D.}~\bibnamefont {Campbell}},
  \bibinfo {author} {\bibfnamefont {C.}~\bibnamefont {Chamon}}, \bibinfo
  {author} {\bibfnamefont {C.-Y.}\ \bibnamefont {Hou}}, \bibinfo {author}
  {\bibfnamefont {R.}~\bibnamefont {Jackiw}}, \bibinfo {author} {\bibfnamefont
  {S.-Y.}\ \bibnamefont {Pi}}, \ and\ \bibinfo {author} {\bibfnamefont {S.~V.}\
  \bibnamefont {Kusminskiy}},\ }\href {\doibase 10.1103/PhysRevLett.110.176603}
  {\bibfield  {journal} {\bibinfo  {journal} {Phys. Rev. Lett.}\ }\textbf
  {\bibinfo {volume} {110}},\ \bibinfo {pages} {176603} (\bibinfo {year}
  {2013})}\BibitemShut {NoStop}%
\bibitem [{\citenamefont {Fregoso}\ \emph {et~al.}(2013)\citenamefont
  {Fregoso}, \citenamefont {Wang}, \citenamefont {Gedik},\ and\ \citenamefont
  {Galitski}}]{Fregoso13}%
  \BibitemOpen
  \bibfield  {author} {\bibinfo {author} {\bibfnamefont {B.~M.}\ \bibnamefont
  {Fregoso}}, \bibinfo {author} {\bibfnamefont {Y.~H.}\ \bibnamefont {Wang}},
  \bibinfo {author} {\bibfnamefont {N.}~\bibnamefont {Gedik}}, \ and\ \bibinfo
  {author} {\bibfnamefont {V.}~\bibnamefont {Galitski}},\ }\href {\doibase
  10.1103/PhysRevB.88.155129} {\bibfield  {journal} {\bibinfo  {journal} {Phys.
  Rev. B}\ }\textbf {\bibinfo {volume} {88}},\ \bibinfo {pages} {155129}
  (\bibinfo {year} {2013})}\BibitemShut {NoStop}%
\bibitem [{\citenamefont {Iadecola}\ \emph {et~al.}(2014)\citenamefont
  {Iadecola}, \citenamefont {Neupert},\ and\ \citenamefont
  {Chamon}}]{Iadecola14}%
  \BibitemOpen
  \bibfield  {author} {\bibinfo {author} {\bibfnamefont {T.}~\bibnamefont
  {Iadecola}}, \bibinfo {author} {\bibfnamefont {T.}~\bibnamefont {Neupert}}, \
  and\ \bibinfo {author} {\bibfnamefont {C.}~\bibnamefont {Chamon}},\ }\href
  {\doibase 10.1103/PhysRevB.89.115425} {\bibfield  {journal} {\bibinfo
  {journal} {Phys. Rev. B}\ }\textbf {\bibinfo {volume} {89}},\ \bibinfo
  {pages} {115425} (\bibinfo {year} {2014})}\BibitemShut {NoStop}%
\bibitem [{\citenamefont {Foa~Torres}\ \emph {et~al.}(2014)\citenamefont
  {Foa~Torres}, \citenamefont {Perez-Piskunow}, \citenamefont {Balseiro},\ and\
  \citenamefont {Usaj}}]{FoaTorres14}%
  \BibitemOpen
  \bibfield  {author} {\bibinfo {author} {\bibfnamefont {L.~E.~F.}\
  \bibnamefont {Foa~Torres}}, \bibinfo {author} {\bibfnamefont {P.~M.}\
  \bibnamefont {Perez-Piskunow}}, \bibinfo {author} {\bibfnamefont {C.~A.}\
  \bibnamefont {Balseiro}}, \ and\ \bibinfo {author} {\bibfnamefont
  {G.}~\bibnamefont {Usaj}},\ }\href@noop {} {\bibfield  {journal} {\bibinfo
  {journal} {arXiv:1409.2482v1}\ } (\bibinfo {year} {2014})}\BibitemShut
  {NoStop}%
\bibitem [{\citenamefont {Sedrakyan}\ \emph {et~al.}(2015)\citenamefont
  {Sedrakyan}, \citenamefont {Galitski},\ and\ \citenamefont
  {Kamenev}}]{Sedrakyan15}%
  \BibitemOpen
  \bibfield  {author} {\bibinfo {author} {\bibfnamefont {T.~A.}\ \bibnamefont
  {Sedrakyan}}, \bibinfo {author} {\bibfnamefont {V.~M.}\ \bibnamefont
  {Galitski}}, \ and\ \bibinfo {author} {\bibfnamefont {A.}~\bibnamefont
  {Kamenev}},\ }\href {http://arxiv.org/abs/1506.00721} {\  (\bibinfo {year}
  {2015})},\ \bibinfo {note} {arXiv:1506.00721v2}\BibitemShut {NoStop}%
\bibitem [{\citenamefont {Kitagawa}\ \emph
  {et~al.}(2012{\natexlab{b}})\citenamefont {Kitagawa}, \citenamefont {Broome},
  \citenamefont {Fedrizzi}, \citenamefont {Rudner}, \citenamefont {Berg},
  \citenamefont {Kassal}, \citenamefont {Aspuru-Guzik}, \citenamefont
  {Demler},\ and\ \citenamefont {White}}]{KitagawaNC12}%
  \BibitemOpen
  \bibfield  {author} {\bibinfo {author} {\bibfnamefont {T.}~\bibnamefont
  {Kitagawa}}, \bibinfo {author} {\bibfnamefont {M.~A.}\ \bibnamefont
  {Broome}}, \bibinfo {author} {\bibfnamefont {A.}~\bibnamefont {Fedrizzi}},
  \bibinfo {author} {\bibfnamefont {M.~S.}\ \bibnamefont {Rudner}}, \bibinfo
  {author} {\bibfnamefont {E.}~\bibnamefont {Berg}}, \bibinfo {author}
  {\bibfnamefont {I.}~\bibnamefont {Kassal}}, \bibinfo {author} {\bibfnamefont
  {A.}~\bibnamefont {Aspuru-Guzik}}, \bibinfo {author} {\bibfnamefont
  {E.}~\bibnamefont {Demler}}, \ and\ \bibinfo {author} {\bibfnamefont {A.~G.}\
  \bibnamefont {White}},\ }\href {\doibase doi:10.1038/ncomms1872} {\bibfield
  {journal} {\bibinfo  {journal} {Nature Communications}\ }\textbf {\bibinfo
  {volume} {3}},\ \bibinfo {pages} {882} (\bibinfo {year}
  {2012}{\natexlab{b}})}\BibitemShut {NoStop}%
\bibitem [{\citenamefont {Rechtsman}\ \emph {et~al.}(2013)\citenamefont
  {Rechtsman}, \citenamefont {Zeuner}, \citenamefont {Plotnik}, \citenamefont
  {Lumer}, \citenamefont {Podolsky}, \citenamefont {Dreisow}, \citenamefont
  {Nolte},\ and\ \citenamefont {Szameit}}]{RechtsmanNature}%
  \BibitemOpen
  \bibfield  {author} {\bibinfo {author} {\bibfnamefont {M.~C.}\ \bibnamefont
  {Rechtsman}}, \bibinfo {author} {\bibfnamefont {J.~M.}\ \bibnamefont
  {Zeuner}}, \bibinfo {author} {\bibfnamefont {Y.}~\bibnamefont {Plotnik}},
  \bibinfo {author} {\bibfnamefont {Y.}~\bibnamefont {Lumer}}, \bibinfo
  {author} {\bibfnamefont {D.}~\bibnamefont {Podolsky}}, \bibinfo {author}
  {\bibfnamefont {F.}~\bibnamefont {Dreisow}}, \bibinfo {author} {\bibfnamefont
  {M.}~\bibnamefont {Nolte}, \bibfnamefont {Stefanand~Segev}}, \ and\ \bibinfo
  {author} {\bibfnamefont {A.}~\bibnamefont {Szameit}},\ }\href {\doibase
  doi:10.1038/nature12066} {\bibfield  {journal} {\bibinfo  {journal} {Nature}\
  }\textbf {\bibinfo {volume} {496}},\ \bibinfo {pages} {196} (\bibinfo {year}
  {2013})}\BibitemShut {NoStop}%
\bibitem [{\citenamefont {Struck}\ \emph {et~al.}(2012)\citenamefont {Struck},
  \citenamefont {\"Olschl\"ager}, \citenamefont {Weinberg}, \citenamefont
  {Hauke}, \citenamefont {Simonet}, \citenamefont {Eckardt}, \citenamefont
  {Lewenstein}, \citenamefont {Sengstock},\ and\ \citenamefont
  {Windpassinger}}]{Struck13}%
  \BibitemOpen
  \bibfield  {author} {\bibinfo {author} {\bibfnamefont {J.}~\bibnamefont
  {Struck}}, \bibinfo {author} {\bibfnamefont {C.}~\bibnamefont
  {\"Olschl\"ager}}, \bibinfo {author} {\bibfnamefont {M.}~\bibnamefont
  {Weinberg}}, \bibinfo {author} {\bibfnamefont {P.}~\bibnamefont {Hauke}},
  \bibinfo {author} {\bibfnamefont {J.}~\bibnamefont {Simonet}}, \bibinfo
  {author} {\bibfnamefont {A.}~\bibnamefont {Eckardt}}, \bibinfo {author}
  {\bibfnamefont {M.}~\bibnamefont {Lewenstein}}, \bibinfo {author}
  {\bibfnamefont {K.}~\bibnamefont {Sengstock}}, \ and\ \bibinfo {author}
  {\bibfnamefont {P.}~\bibnamefont {Windpassinger}},\ }\href {\doibase
  10.1103/PhysRevLett.108.225304} {\bibfield  {journal} {\bibinfo  {journal}
  {Phys. Rev. Lett.}\ }\textbf {\bibinfo {volume} {108}},\ \bibinfo {pages}
  {225304} (\bibinfo {year} {2012})}\BibitemShut {NoStop}%
\bibitem [{\citenamefont {Potter}\ \emph {et~al.}(2016)\citenamefont {Potter},
  \citenamefont {Morimoto},\ and\ \citenamefont
  {Vishwanath}}]{FloquetClassification}%
  \BibitemOpen
  \bibfield  {author} {\bibinfo {author} {\bibfnamefont {A.~C.}\ \bibnamefont
  {Potter}}, \bibinfo {author} {\bibfnamefont {T.}~\bibnamefont {Morimoto}}, \
  and\ \bibinfo {author} {\bibfnamefont {A.}~\bibnamefont {Vishwanath}},\
  }\href {\doibase 10.1103/PhysRevX.6.041001} {\bibfield  {journal} {\bibinfo
  {journal} {Phys. Rev. X}\ }\textbf {\bibinfo {volume} {6}},\ \bibinfo {pages}
  {041001} (\bibinfo {year} {2016})}\BibitemShut {NoStop}%
\bibitem [{\citenamefont {Roy}\ and\ \citenamefont
  {Harper}(2017)}]{FloquetPeriodicTable}%
  \BibitemOpen
  \bibfield  {author} {\bibinfo {author} {\bibfnamefont {R.}~\bibnamefont
  {Roy}}\ and\ \bibinfo {author} {\bibfnamefont {F.}~\bibnamefont {Harper}},\
  }\href {\doibase 10.1103/PhysRevB.96.155118} {\bibfield  {journal} {\bibinfo
  {journal} {Phys. Rev. B}\ }\textbf {\bibinfo {volume} {96}},\ \bibinfo
  {pages} {155118} (\bibinfo {year} {2017})}\BibitemShut {NoStop}%
\bibitem [{\citenamefont {Bomantara}\ and\ \citenamefont
  {Gong}(2018{\natexlab{a}})}]{BomantaraPRL18}%
  \BibitemOpen
  \bibfield  {author} {\bibinfo {author} {\bibfnamefont {R.~W.}\ \bibnamefont
  {Bomantara}}\ and\ \bibinfo {author} {\bibfnamefont {J.}~\bibnamefont
  {Gong}},\ }\href {\doibase 10.1103/PhysRevLett.120.230405} {\bibfield
  {journal} {\bibinfo  {journal} {Phys. Rev. Lett.}\ }\textbf {\bibinfo
  {volume} {120}},\ \bibinfo {pages} {230405} (\bibinfo {year}
  {2018}{\natexlab{a}})}\BibitemShut {NoStop}%
\bibitem [{\citenamefont {Bomantara}\ and\ \citenamefont
  {Gong}(2018{\natexlab{b}})}]{BomantaraPRB18}%
  \BibitemOpen
  \bibfield  {author} {\bibinfo {author} {\bibfnamefont {R.~W.}\ \bibnamefont
  {Bomantara}}\ and\ \bibinfo {author} {\bibfnamefont {J.}~\bibnamefont
  {Gong}},\ }\href {\doibase 10.1103/PhysRevB.98.165421} {\bibfield  {journal}
  {\bibinfo  {journal} {Phys. Rev. B}\ }\textbf {\bibinfo {volume} {98}},\
  \bibinfo {pages} {165421} (\bibinfo {year} {2018}{\natexlab{b}})}\BibitemShut
  {NoStop}%
\bibitem [{\citenamefont {Peng}\ and\ \citenamefont
  {Refael}(2018{\natexlab{a}})}]{PengYangPRB18}%
  \BibitemOpen
  \bibfield  {author} {\bibinfo {author} {\bibfnamefont {Y.}~\bibnamefont
  {Peng}}\ and\ \bibinfo {author} {\bibfnamefont {G.}~\bibnamefont {Refael}},\
  }\href {\doibase 10.1103/PhysRevB.98.220509} {\bibfield  {journal} {\bibinfo
  {journal} {Phys. Rev. B}\ }\textbf {\bibinfo {volume} {98}},\ \bibinfo
  {pages} {220509} (\bibinfo {year} {2018}{\natexlab{a}})}\BibitemShut
  {NoStop}%
\bibitem [{\citenamefont {Bauer}\ \emph {et~al.}(2019)\citenamefont {Bauer},
  \citenamefont {Pereg-Barnea}, \citenamefont {Karzig}, \citenamefont {Rieder},
  \citenamefont {Refael}, \citenamefont {Berg},\ and\ \citenamefont
  {Oreg}}]{BauerPRB19}%
  \BibitemOpen
  \bibfield  {author} {\bibinfo {author} {\bibfnamefont {B.}~\bibnamefont
  {Bauer}}, \bibinfo {author} {\bibfnamefont {T.}~\bibnamefont {Pereg-Barnea}},
  \bibinfo {author} {\bibfnamefont {T.}~\bibnamefont {Karzig}}, \bibinfo
  {author} {\bibfnamefont {M.-T.}\ \bibnamefont {Rieder}}, \bibinfo {author}
  {\bibfnamefont {G.}~\bibnamefont {Refael}}, \bibinfo {author} {\bibfnamefont
  {E.}~\bibnamefont {Berg}}, \ and\ \bibinfo {author} {\bibfnamefont
  {Y.}~\bibnamefont {Oreg}},\ }\href {\doibase 10.1103/PhysRevB.100.041102}
  {\bibfield  {journal} {\bibinfo  {journal} {Phys. Rev. B}\ }\textbf {\bibinfo
  {volume} {100}},\ \bibinfo {pages} {041102} (\bibinfo {year}
  {2019})}\BibitemShut {NoStop}%
\bibitem [{\citenamefont {Yan}\ and\ \citenamefont
  {Wang}(2016)}]{PhysRevLett.117.087402}%
  \BibitemOpen
  \bibfield  {author} {\bibinfo {author} {\bibfnamefont {Z.}~\bibnamefont
  {Yan}}\ and\ \bibinfo {author} {\bibfnamefont {Z.}~\bibnamefont {Wang}},\
  }\href {\doibase 10.1103/PhysRevLett.117.087402} {\bibfield  {journal}
  {\bibinfo  {journal} {Phys. Rev. Lett.}\ }\textbf {\bibinfo {volume} {117}},\
  \bibinfo {pages} {087402} (\bibinfo {year} {2016})}\BibitemShut {NoStop}%
\bibitem [{\citenamefont {Yao}\ \emph {et~al.}(2017)\citenamefont {Yao},
  \citenamefont {Yan},\ and\ \citenamefont {Wang}}]{PhysRevB.96.195303}%
  \BibitemOpen
  \bibfield  {author} {\bibinfo {author} {\bibfnamefont {S.}~\bibnamefont
  {Yao}}, \bibinfo {author} {\bibfnamefont {Z.}~\bibnamefont {Yan}}, \ and\
  \bibinfo {author} {\bibfnamefont {Z.}~\bibnamefont {Wang}},\ }\href {\doibase
  10.1103/PhysRevB.96.195303} {\bibfield  {journal} {\bibinfo  {journal} {Phys.
  Rev. B}\ }\textbf {\bibinfo {volume} {96}},\ \bibinfo {pages} {195303}
  (\bibinfo {year} {2017})}\BibitemShut {NoStop}%
\bibitem [{\citenamefont {Mori}\ \emph {et~al.}(2016)\citenamefont {Mori},
  \citenamefont {Kuwahara},\ and\ \citenamefont {Saito}}]{MoriFloquet}%
  \BibitemOpen
  \bibfield  {author} {\bibinfo {author} {\bibfnamefont {T.}~\bibnamefont
  {Mori}}, \bibinfo {author} {\bibfnamefont {T.}~\bibnamefont {Kuwahara}}, \
  and\ \bibinfo {author} {\bibfnamefont {K.}~\bibnamefont {Saito}},\ }\href
  {\doibase 10.1103/PhysRevLett.116.120401} {\bibfield  {journal} {\bibinfo
  {journal} {Phys. Rev. Lett.}\ }\textbf {\bibinfo {volume} {116}},\ \bibinfo
  {pages} {120401} (\bibinfo {year} {2016})}\BibitemShut {NoStop}%
\bibitem [{\citenamefont {Abanin}\ \emph
  {et~al.}(2017{\natexlab{a}})\citenamefont {Abanin}, \citenamefont {De~Roeck},
  \citenamefont {Ho},\ and\ \citenamefont
  {Huveneers}}]{AbaninFloquetPrethermal1}%
  \BibitemOpen
  \bibfield  {author} {\bibinfo {author} {\bibfnamefont {D.}~\bibnamefont
  {Abanin}}, \bibinfo {author} {\bibfnamefont {W.}~\bibnamefont {De~Roeck}},
  \bibinfo {author} {\bibfnamefont {W.~W.}\ \bibnamefont {Ho}}, \ and\ \bibinfo
  {author} {\bibfnamefont {F.}~\bibnamefont {Huveneers}},\ }\href {\doibase
  10.1007/s00220-017-2930-x} {\bibfield  {journal} {\bibinfo  {journal}
  {Communications in Mathematical Physics}\ }\textbf {\bibinfo {volume}
  {354}},\ \bibinfo {pages} {809} (\bibinfo {year}
  {2017}{\natexlab{a}})}\BibitemShut {NoStop}%
\bibitem [{\citenamefont {Abanin}\ \emph
  {et~al.}(2017{\natexlab{b}})\citenamefont {Abanin}, \citenamefont {De~Roeck},
  \citenamefont {Ho},\ and\ \citenamefont
  {Huveneers}}]{AbaninFloquetPrethermal2}%
  \BibitemOpen
  \bibfield  {author} {\bibinfo {author} {\bibfnamefont {D.~A.}\ \bibnamefont
  {Abanin}}, \bibinfo {author} {\bibfnamefont {W.}~\bibnamefont {De~Roeck}},
  \bibinfo {author} {\bibfnamefont {W.~W.}\ \bibnamefont {Ho}}, \ and\ \bibinfo
  {author} {\bibfnamefont {F.~m.~c.}\ \bibnamefont {Huveneers}},\ }\href
  {\doibase 10.1103/PhysRevB.95.014112} {\bibfield  {journal} {\bibinfo
  {journal} {Phys. Rev. B}\ }\textbf {\bibinfo {volume} {95}},\ \bibinfo
  {pages} {014112} (\bibinfo {year} {2017}{\natexlab{b}})}\BibitemShut
  {NoStop}%
\bibitem [{\citenamefont {Ponte}\ \emph {et~al.}(2015)\citenamefont {Ponte},
  \citenamefont {Chandran}, \citenamefont {Papić},\ and\ \citenamefont
  {Abanin}}]{PONTE2015196}%
  \BibitemOpen
  \bibfield  {author} {\bibinfo {author} {\bibfnamefont {P.}~\bibnamefont
  {Ponte}}, \bibinfo {author} {\bibfnamefont {A.}~\bibnamefont {Chandran}},
  \bibinfo {author} {\bibfnamefont {Z.}~\bibnamefont {Papić}}, \ and\ \bibinfo
  {author} {\bibfnamefont {D.~A.}\ \bibnamefont {Abanin}},\ }\href {\doibase
  https://doi.org/10.1016/j.aop.2014.11.008} {\bibfield  {journal} {\bibinfo
  {journal} {Annals of Physics}\ }\textbf {\bibinfo {volume} {353}},\ \bibinfo
  {pages} {196 } (\bibinfo {year} {2015})}\BibitemShut {NoStop}%
\bibitem [{\citenamefont {Hone}\ \emph {et~al.}(2009)\citenamefont {Hone},
  \citenamefont {Ketzmerick},\ and\ \citenamefont {Kohn}}]{Hone09}%
  \BibitemOpen
  \bibfield  {author} {\bibinfo {author} {\bibfnamefont {D.~W.}\ \bibnamefont
  {Hone}}, \bibinfo {author} {\bibfnamefont {R.}~\bibnamefont {Ketzmerick}}, \
  and\ \bibinfo {author} {\bibfnamefont {W.}~\bibnamefont {Kohn}},\ }\href
  {\doibase 10.1103/PhysRevE.79.051129} {\bibfield  {journal} {\bibinfo
  {journal} {Phys. Rev. E}\ }\textbf {\bibinfo {volume} {79}},\ \bibinfo
  {pages} {051129} (\bibinfo {year} {2009})}\BibitemShut {NoStop}%
\bibitem [{\citenamefont {Liu}(2015)}]{PhysRevB.91.144301}%
  \BibitemOpen
  \bibfield  {author} {\bibinfo {author} {\bibfnamefont {D.~E.}\ \bibnamefont
  {Liu}},\ }\href {\doibase 10.1103/PhysRevB.91.144301} {\bibfield  {journal}
  {\bibinfo  {journal} {Phys. Rev. B}\ }\textbf {\bibinfo {volume} {91}},\
  \bibinfo {pages} {144301} (\bibinfo {year} {2015})}\BibitemShut {NoStop}%
\bibitem [{\citenamefont {Seetharam}\ \emph {et~al.}(2015)\citenamefont
  {Seetharam}, \citenamefont {Bardyn}, \citenamefont {Lindner}, \citenamefont
  {Rudner},\ and\ \citenamefont {Refael}}]{PhysRevX.5.041050}%
  \BibitemOpen
  \bibfield  {author} {\bibinfo {author} {\bibfnamefont {K.~I.}\ \bibnamefont
  {Seetharam}}, \bibinfo {author} {\bibfnamefont {C.-E.}\ \bibnamefont
  {Bardyn}}, \bibinfo {author} {\bibfnamefont {N.~H.}\ \bibnamefont {Lindner}},
  \bibinfo {author} {\bibfnamefont {M.~S.}\ \bibnamefont {Rudner}}, \ and\
  \bibinfo {author} {\bibfnamefont {G.}~\bibnamefont {Refael}},\ }\href
  {\doibase 10.1103/PhysRevX.5.041050} {\bibfield  {journal} {\bibinfo
  {journal} {Phys. Rev. X}\ }\textbf {\bibinfo {volume} {5}},\ \bibinfo {pages}
  {041050} (\bibinfo {year} {2015})}\BibitemShut {NoStop}%
\bibitem [{\citenamefont {Iadecola}\ \emph {et~al.}(2015)\citenamefont
  {Iadecola}, \citenamefont {Neupert},\ and\ \citenamefont
  {Chamon}}]{PhysRevB.91.235133}%
  \BibitemOpen
  \bibfield  {author} {\bibinfo {author} {\bibfnamefont {T.}~\bibnamefont
  {Iadecola}}, \bibinfo {author} {\bibfnamefont {T.}~\bibnamefont {Neupert}}, \
  and\ \bibinfo {author} {\bibfnamefont {C.}~\bibnamefont {Chamon}},\ }\href
  {\doibase 10.1103/PhysRevB.91.235133} {\bibfield  {journal} {\bibinfo
  {journal} {Phys. Rev. B}\ }\textbf {\bibinfo {volume} {91}},\ \bibinfo
  {pages} {235133} (\bibinfo {year} {2015})}\BibitemShut {NoStop}%
\bibitem [{\citenamefont {Fu}\ and\ \citenamefont {Kane}(2008)}]{Fu&Kane08}%
  \BibitemOpen
  \bibfield  {author} {\bibinfo {author} {\bibfnamefont {L.}~\bibnamefont
  {Fu}}\ and\ \bibinfo {author} {\bibfnamefont {C.~L.}\ \bibnamefont {Kane}},\
  }\href {\doibase 10.1103/PhysRevLett.100.096407} {\bibfield  {journal}
  {\bibinfo  {journal} {Phys.\ Rev.\ Lett.}\ }\textbf {\bibinfo {volume}
  {100}},\ \bibinfo {pages} {096407} (\bibinfo {year} {2008})}\BibitemShut
  {NoStop}%
\bibitem [{\citenamefont {Sato}\ \emph {et~al.}(2009)\citenamefont {Sato},
  \citenamefont {Takahashi},\ and\ \citenamefont {Fujimoto}}]{SatoPRL09}%
  \BibitemOpen
  \bibfield  {author} {\bibinfo {author} {\bibfnamefont {M.}~\bibnamefont
  {Sato}}, \bibinfo {author} {\bibfnamefont {Y.}~\bibnamefont {Takahashi}}, \
  and\ \bibinfo {author} {\bibfnamefont {S.}~\bibnamefont {Fujimoto}},\
  }\href@noop {} {\bibfield  {journal} {\bibinfo  {journal} {Phys.\ Rev.\
  Lett.}\ }\textbf {\bibinfo {volume} {103}},\ \bibinfo {pages} {020401}
  (\bibinfo {year} {2009})}\BibitemShut {NoStop}%
\bibitem [{\citenamefont {Sau}\ \emph {et~al.}(2010)\citenamefont {Sau},
  \citenamefont {Lutchyn}, \citenamefont {Tewari},\ and\ \citenamefont {{Das
  Sarma}}}]{Sau10}%
  \BibitemOpen
  \bibfield  {author} {\bibinfo {author} {\bibfnamefont {J.~D.}\ \bibnamefont
  {Sau}}, \bibinfo {author} {\bibfnamefont {R.~M.}\ \bibnamefont {Lutchyn}},
  \bibinfo {author} {\bibfnamefont {S.}~\bibnamefont {Tewari}}, \ and\ \bibinfo
  {author} {\bibfnamefont {S.}~\bibnamefont {{Das Sarma}}},\ }\href {\doibase
  10.1103/PhysRevLett.104.040502} {\bibfield  {journal} {\bibinfo  {journal}
  {Phys.\ Rev.\ Lett.}\ }\textbf {\bibinfo {volume} {104}},\ \bibinfo {pages}
  {040502} (\bibinfo {year} {2010})}\BibitemShut {NoStop}%
\bibitem [{\citenamefont {Lutchyn}\ \emph {et~al.}(2010)\citenamefont
  {Lutchyn}, \citenamefont {Sau},\ and\ \citenamefont
  {Das~Sarma}}]{LutchynPRL10}%
  \BibitemOpen
  \bibfield  {author} {\bibinfo {author} {\bibfnamefont {R.~M.}\ \bibnamefont
  {Lutchyn}}, \bibinfo {author} {\bibfnamefont {J.~D.}\ \bibnamefont {Sau}}, \
  and\ \bibinfo {author} {\bibfnamefont {S.}~\bibnamefont {Das~Sarma}},\ }\href
  {\doibase 10.1103/PhysRevLett.105.077001} {\bibfield  {journal} {\bibinfo
  {journal} {Phys.\ Rev.\ Lett.}\ }\textbf {\bibinfo {volume} {105}},\ \bibinfo
  {pages} {077001} (\bibinfo {year} {2010})}\BibitemShut {NoStop}%
\bibitem [{\citenamefont {Oreg}\ \emph {et~al.}(2010)\citenamefont {Oreg},
  \citenamefont {Refael},\ and\ \citenamefont {von Oppen}}]{1DwiresOreg}%
  \BibitemOpen
  \bibfield  {author} {\bibinfo {author} {\bibfnamefont {Y.}~\bibnamefont
  {Oreg}}, \bibinfo {author} {\bibfnamefont {G.}~\bibnamefont {Refael}}, \ and\
  \bibinfo {author} {\bibfnamefont {F.}~\bibnamefont {von Oppen}},\ }\href
  {\doibase 10.1103/PhysRevLett.105.177002} {\bibfield  {journal} {\bibinfo
  {journal} {Phys.\ Rev.\ Lett.}\ }\textbf {\bibinfo {volume} {105}},\ \bibinfo
  {pages} {177002} (\bibinfo {year} {2010})}\BibitemShut {NoStop}%
\bibitem [{\citenamefont {Stanescu}\ and\ \citenamefont
  {Das~Sarma}(2017)}]{PhysRevB.96.014510}%
  \BibitemOpen
  \bibfield  {author} {\bibinfo {author} {\bibfnamefont {T.~D.}\ \bibnamefont
  {Stanescu}}\ and\ \bibinfo {author} {\bibfnamefont {S.}~\bibnamefont
  {Das~Sarma}},\ }\href {\doibase 10.1103/PhysRevB.96.014510} {\bibfield
  {journal} {\bibinfo  {journal} {Phys. Rev. B}\ }\textbf {\bibinfo {volume}
  {96}},\ \bibinfo {pages} {014510} (\bibinfo {year} {2017})}\BibitemShut
  {NoStop}%
\bibitem [{\citenamefont {Liu}\ \emph {et~al.}(2017)\citenamefont {Liu},
  \citenamefont {Levchenko},\ and\ \citenamefont {Lutchyn}}]{DELPRB17}%
  \BibitemOpen
  \bibfield  {author} {\bibinfo {author} {\bibfnamefont {D.~E.}\ \bibnamefont
  {Liu}}, \bibinfo {author} {\bibfnamefont {A.}~\bibnamefont {Levchenko}}, \
  and\ \bibinfo {author} {\bibfnamefont {R.~M.}\ \bibnamefont {Lutchyn}},\
  }\href {\doibase 10.1103/PhysRevB.95.115303} {\bibfield  {journal} {\bibinfo
  {journal} {Phys. Rev. B}\ }\textbf {\bibinfo {volume} {95}},\ \bibinfo
  {pages} {115303} (\bibinfo {year} {2017})}\BibitemShut {NoStop}%
\bibitem [{\citenamefont {Yates}\ \emph {et~al.}(2018)\citenamefont {Yates},
  \citenamefont {Lemonik},\ and\ \citenamefont
  {Mitra}}]{PhysRevLett.121.076802}%
  \BibitemOpen
  \bibfield  {author} {\bibinfo {author} {\bibfnamefont {D.}~\bibnamefont
  {Yates}}, \bibinfo {author} {\bibfnamefont {Y.}~\bibnamefont {Lemonik}}, \
  and\ \bibinfo {author} {\bibfnamefont {A.}~\bibnamefont {Mitra}},\ }\href
  {\doibase 10.1103/PhysRevLett.121.076802} {\bibfield  {journal} {\bibinfo
  {journal} {Phys. Rev. Lett.}\ }\textbf {\bibinfo {volume} {121}},\ \bibinfo
  {pages} {076802} (\bibinfo {year} {2018})}\BibitemShut {NoStop}%
\bibitem [{\citenamefont {Liu}\ \emph {et~al.}(2019)\citenamefont {Liu},
  \citenamefont {Shabani},\ and\ \citenamefont {Mitra}}]{PhysRevB.99.094303}%
  \BibitemOpen
  \bibfield  {author} {\bibinfo {author} {\bibfnamefont {D.~T.}\ \bibnamefont
  {Liu}}, \bibinfo {author} {\bibfnamefont {J.}~\bibnamefont {Shabani}}, \ and\
  \bibinfo {author} {\bibfnamefont {A.}~\bibnamefont {Mitra}},\ }\href
  {\doibase 10.1103/PhysRevB.99.094303} {\bibfield  {journal} {\bibinfo
  {journal} {Phys. Rev. B}\ }\textbf {\bibinfo {volume} {99}},\ \bibinfo
  {pages} {094303} (\bibinfo {year} {2019})}\BibitemShut {NoStop}%
\bibitem [{\citenamefont {Wyatt}\ \emph {et~al.}(1966)\citenamefont {Wyatt},
  \citenamefont {Dmitriev}, \citenamefont {Moore},\ and\ \citenamefont
  {Sheard}}]{Wyatt66}%
  \BibitemOpen
  \bibfield  {author} {\bibinfo {author} {\bibfnamefont {A.~F.~G.}\
  \bibnamefont {Wyatt}}, \bibinfo {author} {\bibfnamefont {V.~M.}\ \bibnamefont
  {Dmitriev}}, \bibinfo {author} {\bibfnamefont {W.~S.}\ \bibnamefont {Moore}},
  \ and\ \bibinfo {author} {\bibfnamefont {F.~W.}\ \bibnamefont {Sheard}},\
  }\href {\doibase 10.1103/PhysRevLett.16.1166} {\bibfield  {journal} {\bibinfo
   {journal} {Phys. Rev. Lett.}\ }\textbf {\bibinfo {volume} {16}},\ \bibinfo
  {pages} {1166} (\bibinfo {year} {1966})}\BibitemShut {NoStop}%
\bibitem [{\citenamefont {Eliashberg}(1970)}]{Eliashberg70}%
  \BibitemOpen
  \bibfield  {author} {\bibinfo {author} {\bibfnamefont {G.~M.}\ \bibnamefont
  {Eliashberg}},\ }\href
  {http://www.jetpletters.ac.ru/ps/1716/article_26086.pdf} {\bibfield
  {journal} {\bibinfo  {journal} {JETP Letter}\ }\textbf {\bibinfo {volume}
  {11}},\ \bibinfo {pages} {114} (\bibinfo {year} {1970})},\ \bibinfo {note}
  {[Pis'ma Zh. Eksp. Teor. Fiz. 11, 186 (1970)]}\BibitemShut {NoStop}%
\bibitem [{\citenamefont {Robertson}\ and\ \citenamefont
  {Galitski}(2009)}]{Robertson09}%
  \BibitemOpen
  \bibfield  {author} {\bibinfo {author} {\bibfnamefont {A.}~\bibnamefont
  {Robertson}}\ and\ \bibinfo {author} {\bibfnamefont {V.~M.}\ \bibnamefont
  {Galitski}},\ }\href {\doibase 10.1103/PhysRevA.80.063609} {\bibfield
  {journal} {\bibinfo  {journal} {Phys. Rev. A}\ }\textbf {\bibinfo {volume}
  {80}},\ \bibinfo {pages} {063609} (\bibinfo {year} {2009})}\BibitemShut
  {NoStop}%
\bibitem [{\citenamefont {Mankowsky}\ \emph {et~al.}(2014)\citenamefont
  {Mankowsky}, \citenamefont {Subedi}, \citenamefont {Forst}, \citenamefont
  {Mariager}, \citenamefont {Lemke}, \citenamefont {Robinson}, \citenamefont
  {Glownia}, \citenamefont {Minitti}, \citenamefont {Frano}, \citenamefont
  {Fechner}, \citenamefont {Spaldin}, \citenamefont {Loew}, \citenamefont
  {Keimer}, \citenamefont {Georges},\ and\ \citenamefont
  {Cavalleri}}]{Mankowsky14}%
  \BibitemOpen
  \bibfield  {author} {\bibinfo {author} {\bibfnamefont {R.}~\bibnamefont
  {Mankowsky}}, \bibinfo {author} {\bibfnamefont {A.}~\bibnamefont {Subedi}},
  \bibinfo {author} {\bibfnamefont {M.}~\bibnamefont {Forst}}, \bibinfo
  {author} {\bibfnamefont {M.}~\bibnamefont {Mariager}, \bibfnamefont
  {S.~O.~Chollet}}, \bibinfo {author} {\bibfnamefont {H.~T.}\ \bibnamefont
  {Lemke}}, \bibinfo {author} {\bibfnamefont {J.~S.}\ \bibnamefont {Robinson}},
  \bibinfo {author} {\bibfnamefont {J.~M.}\ \bibnamefont {Glownia}}, \bibinfo
  {author} {\bibfnamefont {M.~P.}\ \bibnamefont {Minitti}}, \bibinfo {author}
  {\bibfnamefont {A.}~\bibnamefont {Frano}}, \bibinfo {author} {\bibfnamefont
  {M.}~\bibnamefont {Fechner}}, \bibinfo {author} {\bibfnamefont {N.~A.}\
  \bibnamefont {Spaldin}}, \bibinfo {author} {\bibfnamefont {T.}~\bibnamefont
  {Loew}}, \bibinfo {author} {\bibfnamefont {B.}~\bibnamefont {Keimer}},
  \bibinfo {author} {\bibfnamefont {A.}~\bibnamefont {Georges}}, \ and\
  \bibinfo {author} {\bibfnamefont {A.}~\bibnamefont {Cavalleri}},\ }\href
  {\doibase doi:10.1038/nature13875} {\bibfield  {journal} {\bibinfo  {journal}
  {Nature}\ }\textbf {\bibinfo {volume} {516}},\ \bibinfo {pages} {71}
  (\bibinfo {year} {2014})}\BibitemShut {NoStop}%
\bibitem [{\citenamefont {Galitskii}\ \emph {et~al.}(1969)\citenamefont
  {Galitskii}, \citenamefont {Goreslavskii},\ and\ \citenamefont
  {Elesin}}]{Galitskii69}%
  \BibitemOpen
  \bibfield  {author} {\bibinfo {author} {\bibfnamefont {V.~M.}\ \bibnamefont
  {Galitskii}}, \bibinfo {author} {\bibfnamefont {S.~P.}\ \bibnamefont
  {Goreslavskii}}, \ and\ \bibinfo {author} {\bibfnamefont {V.~F.}\
  \bibnamefont {Elesin}},\ }\href
  {http://jetp.ac.ru/cgi-bin/dn/e_030_01_0117.pdf} {\bibfield  {journal}
  {\bibinfo  {journal} {Sov. Phys. JETP}\ }\textbf {\bibinfo {volume} {30}},\
  \bibinfo {pages} {117} (\bibinfo {year} {1969})},\ \bibinfo {note} {[Zh.
  Eksp. Teor. Fiz. 57, 207 (1969)]}\BibitemShut {NoStop}%
\bibitem [{\citenamefont {Elesin}(1971)}]{Elesin71}%
  \BibitemOpen
  \bibfield  {author} {\bibinfo {author} {\bibfnamefont {V.~F.}\ \bibnamefont
  {Elesin}},\ }\href {http://jetp.ac.ru/cgi-bin/dn/e_032_02_0328.pdf}
  {\bibfield  {journal} {\bibinfo  {journal} {Sov. Phys. JETP}\ }\textbf
  {\bibinfo {volume} {32}},\ \bibinfo {pages} {328} (\bibinfo {year} {1971})},\
  \bibinfo {note} {[Zh. Eksp. Teor. Fiz. 59, 602-614 (1970)]}\BibitemShut
  {NoStop}%
\bibitem [{\citenamefont {Galitskii}\ \emph {et~al.}(1973)\citenamefont
  {Galitskii}, \citenamefont {Elesin},\ and\ \citenamefont
  {Kopaev}}]{Galitskii73}%
  \BibitemOpen
  \bibfield  {author} {\bibinfo {author} {\bibfnamefont {V.~M.}\ \bibnamefont
  {Galitskii}}, \bibinfo {author} {\bibfnamefont {V.~F.}\ \bibnamefont
  {Elesin}}, \ and\ \bibinfo {author} {\bibfnamefont {Y.~V.}\ \bibnamefont
  {Kopaev}},\ }\href {http://www.jetpletters.ac.ru/ps/1539/article_23540.pdf}
  {\bibfield  {journal} {\bibinfo  {journal} {ZhETF Pisma Redaktsiiu}\ }\textbf
  {\bibinfo {volume} {18}},\ \bibinfo {pages} {50} (\bibinfo {year}
  {1973})}\BibitemShut {NoStop}%
\bibitem [{\citenamefont {Elesin}\ \emph {et~al.}(1973)\citenamefont {Elesin},
  \citenamefont {Kopaev},\ and\ \citenamefont {Timerov}}]{Elesin73}%
  \BibitemOpen
  \bibfield  {author} {\bibinfo {author} {\bibfnamefont {V.~F.}\ \bibnamefont
  {Elesin}}, \bibinfo {author} {\bibfnamefont {Y.~V.}\ \bibnamefont {Kopaev}},
  \ and\ \bibinfo {author} {\bibfnamefont {R.~K.}\ \bibnamefont {Timerov}},\
  }\href {http://jetp.ac.ru/cgi-bin/dn/e_038_06_1170.pdf} {\bibfield  {journal}
  {\bibinfo  {journal} {Zh. Eksp. Teor. Fiz.}\ }\textbf {\bibinfo {volume}
  {65}},\ \bibinfo {pages} {2343} (\bibinfo {year} {1973})}\BibitemShut
  {NoStop}%
\bibitem [{\citenamefont {Mani}\ \emph {et~al.}(2002)\citenamefont {Mani},
  \citenamefont {Smet}, \citenamefont {von Klitzing}, \citenamefont
  {Narayanamurti}, \citenamefont {B.},\ and\ \citenamefont
  {Umansky}}]{ManiNature02}%
  \BibitemOpen
  \bibfield  {author} {\bibinfo {author} {\bibfnamefont {R.~G.}\ \bibnamefont
  {Mani}}, \bibinfo {author} {\bibfnamefont {J.~H.}\ \bibnamefont {Smet}},
  \bibinfo {author} {\bibfnamefont {K.}~\bibnamefont {von Klitzing}}, \bibinfo
  {author} {\bibfnamefont {V.}~\bibnamefont {Narayanamurti}}, \bibinfo {author}
  {\bibfnamefont {J.~W.}\ \bibnamefont {B.}}, \ and\ \bibinfo {author}
  {\bibfnamefont {V.}~\bibnamefont {Umansky}},\ }\href {\doibase
  doi:10.1038/nature01277} {\bibfield  {journal} {\bibinfo  {journal} {Nature}\
  }\textbf {\bibinfo {volume} {420}},\ \bibinfo {pages} {646} (\bibinfo {year}
  {2002})}\BibitemShut {NoStop}%
\bibitem [{\citenamefont {Zudov}\ \emph {et~al.}(2003)\citenamefont {Zudov},
  \citenamefont {Du}, \citenamefont {Pfeiffer},\ and\ \citenamefont
  {West}}]{ZudovPRL03}%
  \BibitemOpen
  \bibfield  {author} {\bibinfo {author} {\bibfnamefont {M.~A.}\ \bibnamefont
  {Zudov}}, \bibinfo {author} {\bibfnamefont {R.~R.}\ \bibnamefont {Du}},
  \bibinfo {author} {\bibfnamefont {L.~N.}\ \bibnamefont {Pfeiffer}}, \ and\
  \bibinfo {author} {\bibfnamefont {K.~W.}\ \bibnamefont {West}},\ }\href
  {\doibase 10.1103/PhysRevLett.90.046807} {\bibfield  {journal} {\bibinfo
  {journal} {Phys. Rev. Lett.}\ }\textbf {\bibinfo {volume} {90}},\ \bibinfo
  {pages} {046807} (\bibinfo {year} {2003})}\BibitemShut {NoStop}%
\bibitem [{\citenamefont {Yang}\ \emph {et~al.}(2003)\citenamefont {Yang},
  \citenamefont {Zudov}, \citenamefont {Knuuttila}, \citenamefont {Du},
  \citenamefont {Pfeiffer},\ and\ \citenamefont {West}}]{YangPRL03}%
  \BibitemOpen
  \bibfield  {author} {\bibinfo {author} {\bibfnamefont {C.~L.}\ \bibnamefont
  {Yang}}, \bibinfo {author} {\bibfnamefont {M.~A.}\ \bibnamefont {Zudov}},
  \bibinfo {author} {\bibfnamefont {T.~A.}\ \bibnamefont {Knuuttila}}, \bibinfo
  {author} {\bibfnamefont {R.~R.}\ \bibnamefont {Du}}, \bibinfo {author}
  {\bibfnamefont {L.~N.}\ \bibnamefont {Pfeiffer}}, \ and\ \bibinfo {author}
  {\bibfnamefont {K.~W.}\ \bibnamefont {West}},\ }\href {\doibase
  10.1103/PhysRevLett.91.096803} {\bibfield  {journal} {\bibinfo  {journal}
  {Phys. Rev. Lett.}\ }\textbf {\bibinfo {volume} {91}},\ \bibinfo {pages}
  {096803} (\bibinfo {year} {2003})}\BibitemShut {NoStop}%
\bibitem [{\citenamefont {Andreev}\ \emph {et~al.}(2003)\citenamefont
  {Andreev}, \citenamefont {Aleiner},\ and\ \citenamefont
  {Millis}}]{AndreevPRL03}%
  \BibitemOpen
  \bibfield  {author} {\bibinfo {author} {\bibfnamefont {A.~V.}\ \bibnamefont
  {Andreev}}, \bibinfo {author} {\bibfnamefont {I.~L.}\ \bibnamefont
  {Aleiner}}, \ and\ \bibinfo {author} {\bibfnamefont {A.~J.}\ \bibnamefont
  {Millis}},\ }\href {\doibase 10.1103/PhysRevLett.91.056803} {\bibfield
  {journal} {\bibinfo  {journal} {Phys. Rev. Lett.}\ }\textbf {\bibinfo
  {volume} {91}},\ \bibinfo {pages} {056803} (\bibinfo {year}
  {2003})}\BibitemShut {NoStop}%
\bibitem [{\citenamefont {Vavilov}\ and\ \citenamefont
  {Aleiner}(2004)}]{Vavilov04}%
  \BibitemOpen
  \bibfield  {author} {\bibinfo {author} {\bibfnamefont {M.~G.}\ \bibnamefont
  {Vavilov}}\ and\ \bibinfo {author} {\bibfnamefont {I.~L.}\ \bibnamefont
  {Aleiner}},\ }\href {\doibase 10.1103/PhysRevB.69.035303} {\bibfield
  {journal} {\bibinfo  {journal} {Phys. Rev. B}\ }\textbf {\bibinfo {volume}
  {69}},\ \bibinfo {pages} {035303} (\bibinfo {year} {2004})}\BibitemShut
  {NoStop}%
\bibitem [{\citenamefont {Finkler}\ and\ \citenamefont
  {Halperin}(2009)}]{Finkler09}%
  \BibitemOpen
  \bibfield  {author} {\bibinfo {author} {\bibfnamefont {I.~G.}\ \bibnamefont
  {Finkler}}\ and\ \bibinfo {author} {\bibfnamefont {B.~I.}\ \bibnamefont
  {Halperin}},\ }\href {\doibase 10.1103/PhysRevB.79.085315} {\bibfield
  {journal} {\bibinfo  {journal} {Phys. Rev. B}\ }\textbf {\bibinfo {volume}
  {79}},\ \bibinfo {pages} {085315} (\bibinfo {year} {2009})}\BibitemShut
  {NoStop}%
\bibitem [{\citenamefont {Goldstein}\ \emph {et~al.}(2015)\citenamefont
  {Goldstein}, \citenamefont {Aron},\ and\ \citenamefont
  {Chamon}}]{GoldsteinPRB15}%
  \BibitemOpen
  \bibfield  {author} {\bibinfo {author} {\bibfnamefont {G.}~\bibnamefont
  {Goldstein}}, \bibinfo {author} {\bibfnamefont {C.}~\bibnamefont {Aron}}, \
  and\ \bibinfo {author} {\bibfnamefont {C.}~\bibnamefont {Chamon}},\ }\href
  {\doibase 10.1103/PhysRevB.91.054517} {\bibfield  {journal} {\bibinfo
  {journal} {Phys. Rev. B}\ }\textbf {\bibinfo {volume} {91}},\ \bibinfo
  {pages} {054517} (\bibinfo {year} {2015})}\BibitemShut {NoStop}%
\bibitem [{\citenamefont {Torre}\ \emph {et~al.}(2013)\citenamefont {Torre},
  \citenamefont {Diehl}, \citenamefont {Lukin}, \citenamefont {Sachdev},\ and\
  \citenamefont {Strack}}]{TorrePRA13}%
  \BibitemOpen
  \bibfield  {author} {\bibinfo {author} {\bibfnamefont {E.~G.~D.}\
  \bibnamefont {Torre}}, \bibinfo {author} {\bibfnamefont {S.}~\bibnamefont
  {Diehl}}, \bibinfo {author} {\bibfnamefont {M.~D.}\ \bibnamefont {Lukin}},
  \bibinfo {author} {\bibfnamefont {S.}~\bibnamefont {Sachdev}}, \ and\
  \bibinfo {author} {\bibfnamefont {P.}~\bibnamefont {Strack}},\ }\href
  {\doibase 10.1103/PhysRevA.87.023831} {\bibfield  {journal} {\bibinfo
  {journal} {Phys. Rev. A}\ }\textbf {\bibinfo {volume} {87}},\ \bibinfo
  {pages} {023831} (\bibinfo {year} {2013})}\BibitemShut {NoStop}%
\bibitem [{\citenamefont {Sieberer}\ \emph {et~al.}(2013)\citenamefont
  {Sieberer}, \citenamefont {Huber}, \citenamefont {Altman},\ and\
  \citenamefont {Diehl}}]{SiebererPRL13}%
  \BibitemOpen
  \bibfield  {author} {\bibinfo {author} {\bibfnamefont {L.~M.}\ \bibnamefont
  {Sieberer}}, \bibinfo {author} {\bibfnamefont {S.~D.}\ \bibnamefont {Huber}},
  \bibinfo {author} {\bibfnamefont {E.}~\bibnamefont {Altman}}, \ and\ \bibinfo
  {author} {\bibfnamefont {S.}~\bibnamefont {Diehl}},\ }\href {\doibase
  10.1103/PhysRevLett.110.195301} {\bibfield  {journal} {\bibinfo  {journal}
  {Phys. Rev. Lett.}\ }\textbf {\bibinfo {volume} {110}},\ \bibinfo {pages}
  {195301} (\bibinfo {year} {2013})}\BibitemShut {NoStop}%
\bibitem [{\citenamefont {Sieberer}\ \emph {et~al.}(2014)\citenamefont
  {Sieberer}, \citenamefont {Huber}, \citenamefont {Altman},\ and\
  \citenamefont {Diehl}}]{SiebererPRB14}%
  \BibitemOpen
  \bibfield  {author} {\bibinfo {author} {\bibfnamefont {L.~M.}\ \bibnamefont
  {Sieberer}}, \bibinfo {author} {\bibfnamefont {S.~D.}\ \bibnamefont {Huber}},
  \bibinfo {author} {\bibfnamefont {E.}~\bibnamefont {Altman}}, \ and\ \bibinfo
  {author} {\bibfnamefont {S.}~\bibnamefont {Diehl}},\ }\href {\doibase
  10.1103/PhysRevB.89.134310} {\bibfield  {journal} {\bibinfo  {journal} {Phys.
  Rev. B}\ }\textbf {\bibinfo {volume} {89}},\ \bibinfo {pages} {134310}
  (\bibinfo {year} {2014})}\BibitemShut {NoStop}%
\bibitem [{\citenamefont {Altman}\ \emph {et~al.}(2015)\citenamefont {Altman},
  \citenamefont {Sieberer}, \citenamefont {Chen}, \citenamefont {Diehl},\ and\
  \citenamefont {Toner}}]{AltmanPRX15}%
  \BibitemOpen
  \bibfield  {author} {\bibinfo {author} {\bibfnamefont {E.}~\bibnamefont
  {Altman}}, \bibinfo {author} {\bibfnamefont {L.~M.}\ \bibnamefont
  {Sieberer}}, \bibinfo {author} {\bibfnamefont {L.}~\bibnamefont {Chen}},
  \bibinfo {author} {\bibfnamefont {S.}~\bibnamefont {Diehl}}, \ and\ \bibinfo
  {author} {\bibfnamefont {J.}~\bibnamefont {Toner}},\ }\href {\doibase
  10.1103/PhysRevX.5.011017} {\bibfield  {journal} {\bibinfo  {journal} {Phys.
  Rev. X}\ }\textbf {\bibinfo {volume} {5}},\ \bibinfo {pages} {011017}
  (\bibinfo {year} {2015})}\BibitemShut {NoStop}%
\bibitem [{\citenamefont {Sieberer}\ \emph {et~al.}(2015)\citenamefont
  {Sieberer}, \citenamefont {Buchhold},\ and\ \citenamefont
  {Diehl}}]{Sieberer15}%
  \BibitemOpen
  \bibfield  {author} {\bibinfo {author} {\bibfnamefont {L.~M.}\ \bibnamefont
  {Sieberer}}, \bibinfo {author} {\bibfnamefont {M.}~\bibnamefont {Buchhold}},
  \ and\ \bibinfo {author} {\bibfnamefont {S.}~\bibnamefont {Diehl}},\ }\href
  {https://arxiv.org/abs/1512.00637} {\  (\bibinfo {year} {2015})},\ \bibinfo
  {note} {arXiv:1512.00637}\BibitemShut {NoStop}%
\bibitem [{\citenamefont {Maghrebi}\ and\ \citenamefont
  {Gorshkov}(2016)}]{MaghrebiPRB16}%
  \BibitemOpen
  \bibfield  {author} {\bibinfo {author} {\bibfnamefont {M.~F.}\ \bibnamefont
  {Maghrebi}}\ and\ \bibinfo {author} {\bibfnamefont {A.~V.}\ \bibnamefont
  {Gorshkov}},\ }\href {\doibase 10.1103/PhysRevB.93.014307} {\bibfield
  {journal} {\bibinfo  {journal} {Phys. Rev. B}\ }\textbf {\bibinfo {volume}
  {93}},\ \bibinfo {pages} {014307} (\bibinfo {year} {2016})}\BibitemShut
  {NoStop}%
\bibitem [{\citenamefont {{Mourik}}\ \emph {et~al.}(2012)\citenamefont
  {{Mourik}}, \citenamefont {{Zuo}}, \citenamefont {{Frolov}}, \citenamefont
  {{Plissard}}, \citenamefont {{Bakkers}},\ and\ \citenamefont
  {{Kouwenhoven}}}]{Mourik2012}%
  \BibitemOpen
  \bibfield  {author} {\bibinfo {author} {\bibfnamefont {V.}~\bibnamefont
  {{Mourik}}}, \bibinfo {author} {\bibfnamefont {K.}~\bibnamefont {{Zuo}}},
  \bibinfo {author} {\bibfnamefont {S.~M.}\ \bibnamefont {{Frolov}}}, \bibinfo
  {author} {\bibfnamefont {S.~R.}\ \bibnamefont {{Plissard}}}, \bibinfo
  {author} {\bibfnamefont {E.~P.~A.~M.}\ \bibnamefont {{Bakkers}}}, \ and\
  \bibinfo {author} {\bibfnamefont {L.~P.}\ \bibnamefont {{Kouwenhoven}}},\
  }\href {\doibase 10.1126/science.1222360} {\bibfield  {journal} {\bibinfo
  {journal} {Science}\ }\textbf {\bibinfo {volume} {336}},\ \bibinfo {pages}
  {1003} (\bibinfo {year} {2012})}\BibitemShut {NoStop}%
\bibitem [{\citenamefont {{Deng}}\ \emph {et~al.}(2012)\citenamefont {{Deng}},
  \citenamefont {{Yu}}, \citenamefont {{Huang}}, \citenamefont {{Larsson}},
  \citenamefont {{Caroff}},\ and\ \citenamefont {{Xu}}}]{Deng2012}%
  \BibitemOpen
  \bibfield  {author} {\bibinfo {author} {\bibfnamefont {M.~T.}\ \bibnamefont
  {{Deng}}}, \bibinfo {author} {\bibfnamefont {C.~L.}\ \bibnamefont {{Yu}}},
  \bibinfo {author} {\bibfnamefont {G.~Y.}\ \bibnamefont {{Huang}}}, \bibinfo
  {author} {\bibfnamefont {M.}~\bibnamefont {{Larsson}}}, \bibinfo {author}
  {\bibfnamefont {P.}~\bibnamefont {{Caroff}}}, \ and\ \bibinfo {author}
  {\bibfnamefont {H.~Q.}\ \bibnamefont {{Xu}}},\ }\href {\doibase
  10.1021/nl303758w} {\bibfield  {journal} {\bibinfo  {journal} {Nano Lett.}\
  }\textbf {\bibinfo {volume} {12}},\ \bibinfo {pages} {6414} (\bibinfo {year}
  {2012})}\BibitemShut {NoStop}%
\bibitem [{\citenamefont {{Das}}\ \emph {et~al.}(2012)\citenamefont {{Das}},
  \citenamefont {{Ronen}}, \citenamefont {{Most}}, \citenamefont {{Oreg}},
  \citenamefont {{Heiblum}},\ and\ \citenamefont {{Shtrikman}}}]{Das2012}%
  \BibitemOpen
  \bibfield  {author} {\bibinfo {author} {\bibfnamefont {A.}~\bibnamefont
  {{Das}}}, \bibinfo {author} {\bibfnamefont {Y.}~\bibnamefont {{Ronen}}},
  \bibinfo {author} {\bibfnamefont {Y.}~\bibnamefont {{Most}}}, \bibinfo
  {author} {\bibfnamefont {Y.}~\bibnamefont {{Oreg}}}, \bibinfo {author}
  {\bibfnamefont {M.}~\bibnamefont {{Heiblum}}}, \ and\ \bibinfo {author}
  {\bibfnamefont {H.}~\bibnamefont {{Shtrikman}}},\ }\href {\doibase
  10.1038/nphys2479} {\bibfield  {journal} {\bibinfo  {journal} {Nature Phys.}\
  }\textbf {\bibinfo {volume} {8}},\ \bibinfo {pages} {887} (\bibinfo {year}
  {2012})}\BibitemShut {NoStop}%
\bibitem [{\citenamefont {Churchill}\ \emph {et~al.}(2013)\citenamefont
  {Churchill}, \citenamefont {Fatemi}, \citenamefont {Grove-Rasmussen},
  \citenamefont {Deng}, \citenamefont {Caroff}, \citenamefont {Xu},\ and\
  \citenamefont {Marcus}}]{Churchill2013}%
  \BibitemOpen
  \bibfield  {author} {\bibinfo {author} {\bibfnamefont {H.~O.~H.}\
  \bibnamefont {Churchill}}, \bibinfo {author} {\bibfnamefont {V.}~\bibnamefont
  {Fatemi}}, \bibinfo {author} {\bibfnamefont {K.}~\bibnamefont
  {Grove-Rasmussen}}, \bibinfo {author} {\bibfnamefont {M.~T.}\ \bibnamefont
  {Deng}}, \bibinfo {author} {\bibfnamefont {P.}~\bibnamefont {Caroff}},
  \bibinfo {author} {\bibfnamefont {H.~Q.}\ \bibnamefont {Xu}}, \ and\ \bibinfo
  {author} {\bibfnamefont {C.~M.}\ \bibnamefont {Marcus}},\ }\href {\doibase
  10.1103/PhysRevB.87.241401} {\bibfield  {journal} {\bibinfo  {journal} {Phys.
  Rev. B}\ }\textbf {\bibinfo {volume} {87}},\ \bibinfo {pages} {241401}
  (\bibinfo {year} {2013})}\BibitemShut {NoStop}%
\bibitem [{\citenamefont {Finck}\ \emph {et~al.}(2013)\citenamefont {Finck},
  \citenamefont {Van~Harlingen}, \citenamefont {Mohseni}, \citenamefont
  {Jung},\ and\ \citenamefont {Li}}]{Finck2013}%
  \BibitemOpen
  \bibfield  {author} {\bibinfo {author} {\bibfnamefont {A.~D.~K.}\
  \bibnamefont {Finck}}, \bibinfo {author} {\bibfnamefont {D.~J.}\ \bibnamefont
  {Van~Harlingen}}, \bibinfo {author} {\bibfnamefont {P.~K.}\ \bibnamefont
  {Mohseni}}, \bibinfo {author} {\bibfnamefont {K.}~\bibnamefont {Jung}}, \
  and\ \bibinfo {author} {\bibfnamefont {X.}~\bibnamefont {Li}},\ }\href
  {\doibase 10.1103/PhysRevLett.110.126406} {\bibfield  {journal} {\bibinfo
  {journal} {Phys. Rev. Lett.}\ }\textbf {\bibinfo {volume} {110}},\ \bibinfo
  {pages} {126406} (\bibinfo {year} {2013})}\BibitemShut {NoStop}%
\bibitem [{\citenamefont {Albrecht}\ \emph {et~al.}(2016)\citenamefont
  {Albrecht}, \citenamefont {Higginbotham}, \citenamefont {Madsen},
  \citenamefont {Kuemmeth}, \citenamefont {Jespersen}, \citenamefont
  {Krogstrup},\ and\ \citenamefont {Marcus}}]{Albrecht16}%
  \BibitemOpen
  \bibfield  {author} {\bibinfo {author} {\bibfnamefont {S.~M.}\ \bibnamefont
  {Albrecht}}, \bibinfo {author} {\bibfnamefont {A.~P.}\ \bibnamefont
  {Higginbotham}}, \bibinfo {author} {\bibfnamefont {M.}~\bibnamefont
  {Madsen}}, \bibinfo {author} {\bibfnamefont {F.}~\bibnamefont {Kuemmeth}},
  \bibinfo {author} {\bibfnamefont {J.}~\bibnamefont {Jespersen}, \bibfnamefont
  {T.~S.~Nygard}}, \bibinfo {author} {\bibfnamefont {P.}~\bibnamefont
  {Krogstrup}}, \ and\ \bibinfo {author} {\bibfnamefont {C.~M.}\ \bibnamefont
  {Marcus}},\ }\href@noop {} {\bibfield  {journal} {\bibinfo  {journal}
  {Nature}\ }\textbf {\bibinfo {volume} {531}},\ \bibinfo {pages} {206}
  (\bibinfo {year} {2016})}\BibitemShut {NoStop}%
\bibitem [{\citenamefont {Deng}\ \emph {et~al.}(2016)\citenamefont {Deng},
  \citenamefont {Vaitiek{\.e}nas}, \citenamefont {Hansen}, \citenamefont
  {Danon}, \citenamefont {Leijnse}, \citenamefont {Flensberg}, \citenamefont
  {Nyg{\aa}rd}, \citenamefont {Krogstrup},\ and\ \citenamefont
  {Marcus}}]{deng2016Majorana}%
  \BibitemOpen
  \bibfield  {author} {\bibinfo {author} {\bibfnamefont {M.}~\bibnamefont
  {Deng}}, \bibinfo {author} {\bibfnamefont {S.}~\bibnamefont
  {Vaitiek{\.e}nas}}, \bibinfo {author} {\bibfnamefont {E.~B.}\ \bibnamefont
  {Hansen}}, \bibinfo {author} {\bibfnamefont {J.}~\bibnamefont {Danon}},
  \bibinfo {author} {\bibfnamefont {M.}~\bibnamefont {Leijnse}}, \bibinfo
  {author} {\bibfnamefont {K.}~\bibnamefont {Flensberg}}, \bibinfo {author}
  {\bibfnamefont {J.}~\bibnamefont {Nyg{\aa}rd}}, \bibinfo {author}
  {\bibfnamefont {P.}~\bibnamefont {Krogstrup}}, \ and\ \bibinfo {author}
  {\bibfnamefont {C.~M.}\ \bibnamefont {Marcus}},\ }\href@noop {} {\bibfield
  {journal} {\bibinfo  {journal} {Science}\ }\textbf {\bibinfo {volume}
  {354}},\ \bibinfo {pages} {1557} (\bibinfo {year} {2016})}\BibitemShut
  {NoStop}%
\bibitem [{\citenamefont {Zhang}\ \emph {et~al.}(2017)\citenamefont {Zhang},
  \citenamefont {G{\"u}l}, \citenamefont {Conesa-Boj}, \citenamefont {Nowak},
  \citenamefont {Wimmer}, \citenamefont {Zuo}, \citenamefont {Mourik},
  \citenamefont {de~Vries}, \citenamefont {van Veen}, \citenamefont {de~Moor},
  \citenamefont {Bommer}, \citenamefont {van Woerkom}, \citenamefont {Car},
  \citenamefont {Plissard}, \citenamefont {Bakkers}, \citenamefont
  {Quintero-P{\'e}rez}, \citenamefont {Cassidy}, \citenamefont {Koelling},
  \citenamefont {Goswami}, \citenamefont {Watanabe}, \citenamefont
  {Taniguchi},\ and\ \citenamefont {Kouwenhoven}}]{Zhang2017Ballistic}%
  \BibitemOpen
  \bibfield  {author} {\bibinfo {author} {\bibfnamefont {H.}~\bibnamefont
  {Zhang}}, \bibinfo {author} {\bibfnamefont {{\"O}.}~\bibnamefont {G{\"u}l}},
  \bibinfo {author} {\bibfnamefont {S.}~\bibnamefont {Conesa-Boj}}, \bibinfo
  {author} {\bibfnamefont {M.}~\bibnamefont {Nowak}}, \bibinfo {author}
  {\bibfnamefont {M.}~\bibnamefont {Wimmer}}, \bibinfo {author} {\bibfnamefont
  {K.}~\bibnamefont {Zuo}}, \bibinfo {author} {\bibfnamefont {V.}~\bibnamefont
  {Mourik}}, \bibinfo {author} {\bibfnamefont {F.~K.}\ \bibnamefont
  {de~Vries}}, \bibinfo {author} {\bibfnamefont {J.}~\bibnamefont {van Veen}},
  \bibinfo {author} {\bibfnamefont {M.~W.~A.}\ \bibnamefont {de~Moor}},
  \bibinfo {author} {\bibfnamefont {J.~D.~S.}\ \bibnamefont {Bommer}}, \bibinfo
  {author} {\bibfnamefont {D.~J.}\ \bibnamefont {van Woerkom}}, \bibinfo
  {author} {\bibfnamefont {D.}~\bibnamefont {Car}}, \bibinfo {author}
  {\bibfnamefont {S.~R.}\ \bibnamefont {Plissard}}, \bibinfo {author}
  {\bibfnamefont {E.~P. A.~M.}\ \bibnamefont {Bakkers}}, \bibinfo {author}
  {\bibfnamefont {M.}~\bibnamefont {Quintero-P{\'e}rez}}, \bibinfo {author}
  {\bibfnamefont {M.~C.}\ \bibnamefont {Cassidy}}, \bibinfo {author}
  {\bibfnamefont {S.}~\bibnamefont {Koelling}}, \bibinfo {author}
  {\bibfnamefont {S.}~\bibnamefont {Goswami}}, \bibinfo {author} {\bibfnamefont
  {K.}~\bibnamefont {Watanabe}}, \bibinfo {author} {\bibfnamefont
  {T.}~\bibnamefont {Taniguchi}}, \ and\ \bibinfo {author} {\bibfnamefont
  {L.~P.}\ \bibnamefont {Kouwenhoven}},\ }\href
  {https://doi.org/10.1038/ncomms16025} {\bibfield  {journal} {\bibinfo
  {journal} {Nature Communications}\ }\textbf {\bibinfo {volume} {8}},\
  \bibinfo {pages} {16025} (\bibinfo {year} {2017})}\BibitemShut {NoStop}%
\bibitem [{\citenamefont {G{\"u}l}\ \emph {et~al.}(2018)\citenamefont
  {G{\"u}l}, \citenamefont {Zhang}, \citenamefont {Bommer}, \citenamefont
  {de~Moor}, \citenamefont {Car}, \citenamefont {Plissard}, \citenamefont
  {Bakkers}, \citenamefont {Geresdi}, \citenamefont {Watanabe}, \citenamefont
  {Taniguchi},\ and\ \citenamefont {Kouwenhoven}}]{ZhangNN2018}%
  \BibitemOpen
  \bibfield  {author} {\bibinfo {author} {\bibfnamefont {{\"O}.}~\bibnamefont
  {G{\"u}l}}, \bibinfo {author} {\bibfnamefont {H.}~\bibnamefont {Zhang}},
  \bibinfo {author} {\bibfnamefont {J.~D.~S.}\ \bibnamefont {Bommer}}, \bibinfo
  {author} {\bibfnamefont {M.~W.~A.}\ \bibnamefont {de~Moor}}, \bibinfo
  {author} {\bibfnamefont {D.}~\bibnamefont {Car}}, \bibinfo {author}
  {\bibfnamefont {S.~R.}\ \bibnamefont {Plissard}}, \bibinfo {author}
  {\bibfnamefont {E.~P. A.~M.}\ \bibnamefont {Bakkers}}, \bibinfo {author}
  {\bibfnamefont {A.}~\bibnamefont {Geresdi}}, \bibinfo {author} {\bibfnamefont
  {K.}~\bibnamefont {Watanabe}}, \bibinfo {author} {\bibfnamefont
  {T.}~\bibnamefont {Taniguchi}}, \ and\ \bibinfo {author} {\bibfnamefont
  {L.~P.}\ \bibnamefont {Kouwenhoven}},\ }\href {\doibase
  10.1038/s41565-017-0032-8} {\bibfield  {journal} {\bibinfo  {journal} {Nature
  Nanotechnology}\ }\textbf {\bibinfo {volume} {13}},\ \bibinfo {pages} {192}
  (\bibinfo {year} {2018})}\BibitemShut {NoStop}%
\bibitem [{sup()}]{supp}%
  \BibitemOpen
  \href@noop {} {}\bibinfo {note} {See Supplemental Material [url] for (i)
  Derivation of the Floquet Retarded Green's Function; (ii) Recursive Green's
  Function Method; (iii) Topological phase transition of Floquet Majorana zero
  modes; (iv) Some additional notes of the Majorana poisoning model; which
  includes additional
  Refs.~\cite{RevModPhys.86.779,RevModPhys.58.323,ch2003introduction,KOHLER2005379,Brody_2013}}\BibitemShut
  {NoStop}%
\bibitem [{not()}]{note}%
  \BibitemOpen
  \href@noop {} {}\bibinfo {note} {We note that the retarded Green's function
  in Eq. 2 is defined as the Fourier transformation of the following Green's
  function defined in time domain
  $G_{nw}^R(k;t,t')=-i\theta(t-t')\langle\{\hat{\Psi}_k(t),\hat{\Psi}^\dag_k(t')\}\rangle$,
  where
  $\hat{\Psi}_k(t)=(\hat{c}_{k,\uparrow}(t),\hat{c}_{k,\downarrow}(t),\hat{c}_{-k,\uparrow}^\dag(t),\hat{c}_{-k,\downarrow}^\dag(t))^t$
  is the field operator defiend in the Heisenberg picture.}\BibitemShut {Stop}%
\bibitem [{Mah(1981)}]{Mahan}%
  \BibitemOpen
  \href@noop {} {\emph {\bibinfo {title} {Many-Particle Physics}}}\ (\bibinfo
  {year} {1981})\BibitemShut {NoStop}%
\bibitem [{Note1()}]{Note1}%
  \BibitemOpen
  \bibinfo {note} {We note that in the static case with the parameters we
  chosen, the system is topological nontrivial and has Majorana zero modes on
  the boundary.}\BibitemShut {Stop}%
\bibitem [{\citenamefont {Peng}\ and\ \citenamefont
  {Refael}(2018{\natexlab{b}})}]{PhysRevB.98.220509}%
  \BibitemOpen
  \bibfield  {author} {\bibinfo {author} {\bibfnamefont {Y.}~\bibnamefont
  {Peng}}\ and\ \bibinfo {author} {\bibfnamefont {G.}~\bibnamefont {Refael}},\
  }\href {\doibase 10.1103/PhysRevB.98.220509} {\bibfield  {journal} {\bibinfo
  {journal} {Phys. Rev. B}\ }\textbf {\bibinfo {volume} {98}},\ \bibinfo
  {pages} {220509} (\bibinfo {year} {2018}{\natexlab{b}})}\BibitemShut
  {NoStop}%
\bibitem [{Note2()}]{Note2}%
  \BibitemOpen
  \bibinfo {note} {We note that in Fig.~\ref {F2} (a), the parameters $\Delta
  =200$, $N_F=5$, $\Omega =6$ satisfy the condition $\Delta \gg (2N_F+1)\Omega
  $.}\BibitemShut {Stop}%
\bibitem [{\citenamefont {Thouless}\ and\ \citenamefont
  {Kirkpatrick}(1981)}]{Thouless_1981}%
  \BibitemOpen
  \bibfield  {author} {\bibinfo {author} {\bibfnamefont {D.~J.}\ \bibnamefont
  {Thouless}}\ and\ \bibinfo {author} {\bibfnamefont {S.}~\bibnamefont
  {Kirkpatrick}},\ }\href {\doibase 10.1088/0022-3719/14/3/007} {\bibfield
  {journal} {\bibinfo  {journal} {Journal of Physics C: Solid State Physics}\
  }\textbf {\bibinfo {volume} {14}},\ \bibinfo {pages} {235} (\bibinfo {year}
  {1981})}\BibitemShut {NoStop}%
\bibitem [{\citenamefont {Lee}\ and\ \citenamefont
  {Fisher}(1981)}]{PhysRevLett.47.882}%
  \BibitemOpen
  \bibfield  {author} {\bibinfo {author} {\bibfnamefont {P.~A.}\ \bibnamefont
  {Lee}}\ and\ \bibinfo {author} {\bibfnamefont {D.~S.}\ \bibnamefont
  {Fisher}},\ }\href {\doibase 10.1103/PhysRevLett.47.882} {\bibfield
  {journal} {\bibinfo  {journal} {Phys. Rev. Lett.}\ }\textbf {\bibinfo
  {volume} {47}},\ \bibinfo {pages} {882} (\bibinfo {year} {1981})}\BibitemShut
  {NoStop}%
\bibitem [{\citenamefont {Drouvelis}\ \emph {et~al.}(2006)\citenamefont
  {Drouvelis}, \citenamefont {Schmelcher},\ and\ \citenamefont
  {Bastian}}]{DROUVELIS2006741}%
  \BibitemOpen
  \bibfield  {author} {\bibinfo {author} {\bibfnamefont {P.~S.}\ \bibnamefont
  {Drouvelis}}, \bibinfo {author} {\bibfnamefont {P.}~\bibnamefont
  {Schmelcher}}, \ and\ \bibinfo {author} {\bibfnamefont {P.}~\bibnamefont
  {Bastian}},\ }\href {\doibase https://doi.org/10.1016/j.jcp.2005.11.010}
  {\bibfield  {journal} {\bibinfo  {journal} {Journal of Computational
  Physics}\ }\textbf {\bibinfo {volume} {215}},\ \bibinfo {pages} {741 }
  (\bibinfo {year} {2006})}\BibitemShut {NoStop}%
\bibitem [{Note3()}]{Note3}%
  \BibitemOpen
  \bibinfo {note} {Since the Floquet systems only have discrete time
  translational symmetry, the energy is only conserved up to $m\Omega $.
  Therefore, similar to the crystal momentum, one can define the Floquet
  Brillouin zones for the quasi-energies as follows $\omega \in [(m-1/2)\Omega
  ,(m+1/2)\Omega ]$ for $m=-N_F,...,N_F$. When $m=0$, $\omega \in [-\Omega
  /2,\Omega /2]$ defines the so-called first Floquet Brillouin
  zones.}\BibitemShut {Stop}%
\bibitem [{\citenamefont {{Kozii}}\ and\ \citenamefont
  {{Fu}}(2017)}]{2017arXiv170805841K}%
  \BibitemOpen
  \bibfield  {author} {\bibinfo {author} {\bibfnamefont {V.}~\bibnamefont
  {{Kozii}}}\ and\ \bibinfo {author} {\bibfnamefont {L.}~\bibnamefont {{Fu}}},\
  }\href@noop {} {\bibfield  {journal} {\bibinfo  {journal} {arXiv e-prints}\
  ,\ \bibinfo {eid} {arXiv:1708.05841}} (\bibinfo {year} {2017})},\ \Eprint
  {http://arxiv.org/abs/1708.05841} {arXiv:1708.05841 [cond-mat.mes-hall]}
  \BibitemShut {NoStop}%
\bibitem [{\citenamefont {Papaj}\ \emph {et~al.}(2019)\citenamefont {Papaj},
  \citenamefont {Isobe},\ and\ \citenamefont {Fu}}]{PhysRevB.99.201107}%
  \BibitemOpen
  \bibfield  {author} {\bibinfo {author} {\bibfnamefont {M.}~\bibnamefont
  {Papaj}}, \bibinfo {author} {\bibfnamefont {H.}~\bibnamefont {Isobe}}, \ and\
  \bibinfo {author} {\bibfnamefont {L.}~\bibnamefont {Fu}},\ }\href {\doibase
  10.1103/PhysRevB.99.201107} {\bibfield  {journal} {\bibinfo  {journal} {Phys.
  Rev. B}\ }\textbf {\bibinfo {volume} {99}},\ \bibinfo {pages} {201107}
  (\bibinfo {year} {2019})}\BibitemShut {NoStop}%
\bibitem [{\citenamefont {Shen}\ and\ \citenamefont
  {Fu}(2018)}]{PhysRevLett.121.026403}%
  \BibitemOpen
  \bibfield  {author} {\bibinfo {author} {\bibfnamefont {H.}~\bibnamefont
  {Shen}}\ and\ \bibinfo {author} {\bibfnamefont {L.}~\bibnamefont {Fu}},\
  }\href {\doibase 10.1103/PhysRevLett.121.026403} {\bibfield  {journal}
  {\bibinfo  {journal} {Phys. Rev. Lett.}\ }\textbf {\bibinfo {volume} {121}},\
  \bibinfo {pages} {026403} (\bibinfo {year} {2018})}\BibitemShut {NoStop}%
\bibitem [{\citenamefont {Yoshida}\ \emph {et~al.}(2018)\citenamefont
  {Yoshida}, \citenamefont {Peters},\ and\ \citenamefont
  {Kawakami}}]{PhysRevB.98.035141}%
  \BibitemOpen
  \bibfield  {author} {\bibinfo {author} {\bibfnamefont {T.}~\bibnamefont
  {Yoshida}}, \bibinfo {author} {\bibfnamefont {R.}~\bibnamefont {Peters}}, \
  and\ \bibinfo {author} {\bibfnamefont {N.}~\bibnamefont {Kawakami}},\ }\href
  {\doibase 10.1103/PhysRevB.98.035141} {\bibfield  {journal} {\bibinfo
  {journal} {Phys. Rev. B}\ }\textbf {\bibinfo {volume} {98}},\ \bibinfo
  {pages} {035141} (\bibinfo {year} {2018})}\BibitemShut {NoStop}%
\bibitem [{\citenamefont {Moors}\ \emph {et~al.}(2019)\citenamefont {Moors},
  \citenamefont {Zyuzin}, \citenamefont {Zyuzin}, \citenamefont {Tiwari},\ and\
  \citenamefont {Schmidt}}]{PhysRevB.99.041116}%
  \BibitemOpen
  \bibfield  {author} {\bibinfo {author} {\bibfnamefont {K.}~\bibnamefont
  {Moors}}, \bibinfo {author} {\bibfnamefont {A.~A.}\ \bibnamefont {Zyuzin}},
  \bibinfo {author} {\bibfnamefont {A.~Y.}\ \bibnamefont {Zyuzin}}, \bibinfo
  {author} {\bibfnamefont {R.~P.}\ \bibnamefont {Tiwari}}, \ and\ \bibinfo
  {author} {\bibfnamefont {T.~L.}\ \bibnamefont {Schmidt}},\ }\href {\doibase
  10.1103/PhysRevB.99.041116} {\bibfield  {journal} {\bibinfo  {journal} {Phys.
  Rev. B}\ }\textbf {\bibinfo {volume} {99}},\ \bibinfo {pages} {041116}
  (\bibinfo {year} {2019})}\BibitemShut {NoStop}%
\bibitem [{\citenamefont {{Yi}}\ and\ \citenamefont
  {{Yang}}(2020)}]{2020arXiv200302219Y}%
  \BibitemOpen
  \bibfield  {author} {\bibinfo {author} {\bibfnamefont {Y.}~\bibnamefont
  {{Yi}}}\ and\ \bibinfo {author} {\bibfnamefont {Z.}~\bibnamefont {{Yang}}},\
  }\href@noop {} {\bibfield  {journal} {\bibinfo  {journal} {arXiv e-prints}\
  ,\ \bibinfo {eid} {arXiv:2003.02219}} (\bibinfo {year} {2020})},\ \Eprint
  {http://arxiv.org/abs/2003.02219} {arXiv:2003.02219 [cond-mat.mes-hall]}
  \BibitemShut {NoStop}%
\bibitem [{\citenamefont {Aoki}\ \emph {et~al.}(2014)\citenamefont {Aoki},
  \citenamefont {Tsuji}, \citenamefont {Eckstein}, \citenamefont {Kollar},
  \citenamefont {Oka},\ and\ \citenamefont {Werner}}]{RevModPhys.86.779}%
  \BibitemOpen
  \bibfield  {author} {\bibinfo {author} {\bibfnamefont {H.}~\bibnamefont
  {Aoki}}, \bibinfo {author} {\bibfnamefont {N.}~\bibnamefont {Tsuji}},
  \bibinfo {author} {\bibfnamefont {M.}~\bibnamefont {Eckstein}}, \bibinfo
  {author} {\bibfnamefont {M.}~\bibnamefont {Kollar}}, \bibinfo {author}
  {\bibfnamefont {T.}~\bibnamefont {Oka}}, \ and\ \bibinfo {author}
  {\bibfnamefont {P.}~\bibnamefont {Werner}},\ }\href {\doibase
  10.1103/RevModPhys.86.779} {\bibfield  {journal} {\bibinfo  {journal} {Rev.
  Mod. Phys.}\ }\textbf {\bibinfo {volume} {86}},\ \bibinfo {pages} {779}
  (\bibinfo {year} {2014})}\BibitemShut {NoStop}%
\bibitem [{\citenamefont {Rammer}\ and\ \citenamefont
  {Smith}(1986)}]{RevModPhys.58.323}%
  \BibitemOpen
  \bibfield  {author} {\bibinfo {author} {\bibfnamefont {J.}~\bibnamefont
  {Rammer}}\ and\ \bibinfo {author} {\bibfnamefont {H.}~\bibnamefont {Smith}},\
  }\href {\doibase 10.1103/RevModPhys.58.323} {\bibfield  {journal} {\bibinfo
  {journal} {Rev. Mod. Phys.}\ }\textbf {\bibinfo {volume} {58}},\ \bibinfo
  {pages} {323} (\bibinfo {year} {1986})}\BibitemShut {NoStop}%
\bibitem [{\citenamefont {Chandrasekhar}(2003)}]{ch2003introduction}%
  \BibitemOpen
  \bibfield  {author} {\bibinfo {author} {\bibfnamefont {V.}~\bibnamefont
  {Chandrasekhar}},\ }\href@noop {} {\enquote {\bibinfo {title} {An
  introduction to the quasiclassical theory of superconductivity for diffusive
  proximity-coupled systems},}\ } (\bibinfo {year} {2003}),\ \Eprint
  {http://arxiv.org/abs/cond-mat/0312507} {arXiv:cond-mat/0312507
  [cond-mat.mes-hall]} \BibitemShut {NoStop}%
\bibitem [{\citenamefont {Kohler}\ \emph {et~al.}(2005)\citenamefont {Kohler},
  \citenamefont {Lehmann},\ and\ \citenamefont {H{\"a}nggi}}]{KOHLER2005379}%
  \BibitemOpen
  \bibfield  {author} {\bibinfo {author} {\bibfnamefont {S.}~\bibnamefont
  {Kohler}}, \bibinfo {author} {\bibfnamefont {J.}~\bibnamefont {Lehmann}}, \
  and\ \bibinfo {author} {\bibfnamefont {P.}~\bibnamefont {H{\"a}nggi}},\
  }\href {\doibase https://doi.org/10.1016/j.physrep.2004.11.002} {\bibfield
  {journal} {\bibinfo  {journal} {Physics Reports}\ }\textbf {\bibinfo {volume}
  {406}},\ \bibinfo {pages} {379 } (\bibinfo {year} {2005})}\BibitemShut
  {NoStop}%
\bibitem [{\citenamefont {Brody}(2013)}]{Brody_2013}%
  \BibitemOpen
  \bibfield  {author} {\bibinfo {author} {\bibfnamefont {D.~C.}\ \bibnamefont
  {Brody}},\ }\href {\doibase 10.1088/1751-8113/47/3/035305} {\bibfield
  {journal} {\bibinfo  {journal} {Journal of Physics A: Mathematical and
  Theoretical}\ }\textbf {\bibinfo {volume} {47}},\ \bibinfo {pages} {035305}
  (\bibinfo {year} {2013})}\BibitemShut {NoStop}%
\end{thebibliography}%
\bibliographystyle{apsrev4-1}

\onecolumngrid
\newpage


\section*{Supplementary material}

# Supplementary Material for "Dissipative Floquet Majorana modes in proximity-induced topological superconductors"

**CONTENTS**

## I. DERIVATION OF THE FLOQUET RETARDED GREEN'S FUNCTION

In this section, we will briefly show how to derive $G^R_{nw}(k,\omega)$. In order to make the Supplementary Material (SM) self-contained, we write down our Hamiltonian

$$\hat{H}(t) = \hat{H}_{nw}(t) + \hat{H}_{sc} + \hat{H}_c,$$
$$\hat{H}_{nw}(t) = \sum_k \hat{\Psi}^\dagger_k [(-2t_0 \cos k - \mu_0 + 2A\cos\Omega t)\tau_z + V_z \sigma_z \tau_z + \lambda \sin k \sigma_y \tau_z]\hat{\Psi}_k,$$
$$\hat{H}_{sc} = \sum_q \hat{\Phi}^\dagger_q (\epsilon_q \tau_z - \Delta \sigma_y \tau_y)\hat{\Phi}_q, \quad (1)$$
$$\hat{H}_c = \sum_{k,q,\sigma}(V \hat{c}^\dagger_{k,\sigma} \hat{a}_{q,\sigma} + V^* \hat{a}^\dagger_{q,\sigma}\hat{c}_{k,\sigma}),$$

where

$$\hat{\Psi}_k = (\hat{c}_{k,\uparrow}, \hat{c}_{k,\downarrow}, \hat{c}^\dagger_{-k,\uparrow}, \hat{c}^\dagger_{-k,\downarrow})^t, \qquad \hat{\Phi}_q = (\hat{a}_{q,\uparrow}, \hat{a}_{q,\downarrow}, \hat{a}^\dagger_{-q,\uparrow}, \hat{a}^\dagger_{-q,\downarrow})^t \quad (2)$$

are written in the BdG basis; $\hat{c}_{k,\uparrow/\downarrow}$ annihilate electrons with spin up/down and momentum $k$ in the nanowire; $\hat{a}_{q,\uparrow/\downarrow}$ annihilate electrons with spin up/down and momentum $q$ in the superconductivity (SC) bath; $\sigma_\mu$ and $\tau_\mu$ represent the Pauli matrices in the spin and Nambu spaces, respectively. The definition of other parameters can be referred to the main text.

## A. Floquet's theorem

In this subsection, we will use the Floquet theorem to translate the time-dependent problem to a time-independent problem. Using $\hat{\Psi}_k(t)$ and $\hat{H}(t)$ to label the field operators and the Hamiltonian in the Heisenberg picture, the dynamics of $\hat{\Psi}_k(t)$ is determined by the following Heisenberg equation,

$$i\partial_t \hat{\Psi}_k(t) = \left[\hat{\Psi}_k(t), \hat{H}(t)\right]. \tag{3}$$

We here emphasize that when the Hamiltonian $\hat{H}(t)$ is time-dependent, the Hamiltonian in the Schrodinger picture $\hat{H}_S(t)$ is not equal to the one in the Heisenberg picture $\hat{H}(t) := \hat{H}_H(t) = \hat{U}^\dagger(t)\hat{H}_S(t)\hat{U}(t)$, where $\hat{U}(t) = T[e^{-i\int_0^t \hat{H}_S(t')dt'}]$. Since the full Hamiltonian $\hat{H}(t)$ is quadratic, one can apply the following relation in the commutator calculation of Eq. 3

$$[A(t), B(t)C(t)] = \{A(t), B(t)\}C(t) - B(t)\{A(t), C(t)\}, \tag{4}$$

and obtain

$$\begin{aligned}
i\partial_t \hat{\Psi}_k(t) &= \left[\hat{\Psi}_k(t), \hat{H}_{nw}(t) + \hat{H}_{sc} + \hat{H}_c\right] \\
&= \left[\hat{\Psi}_k(t), \hat{H}_{nw}(t) + \hat{H}_c\right] \\
&= \left[\hat{\Psi}_k(t), \sum_k \hat{\Psi}_k^\dagger(t) H_{nw}(k,t)\hat{\Psi}_k(t) + \sum_{k,q} \hat{\Psi}_k^\dagger(t) V\sigma_0\tau_z \hat{\Phi}_q(t) + \sum_{k,q} \hat{\Phi}_q^\dagger(t) V\sigma_0\tau_z \hat{\Psi}_k(t)\right] \\
&= H_{nw}(k,t)\hat{\Psi}_k(t) + \sum_q V\sigma_0\tau_z \hat{\Phi}_q(t)
\end{aligned} \tag{5}$$

where

$$H_{nw}(k,t) = [(-2t_0 \cos k - \mu_0 + 2A\cos\Omega t)\sigma_0 + V_z\sigma_z + \lambda \sin k \sigma_y]\tau_z, \tag{6}$$

is the time dependent first quantized Hamiltonian of the nanowire. Following the same procedure, one can obtain

$$i\frac{\partial}{\partial t} \begin{pmatrix} \hat{\Psi}_k(t) \\ \hat{\Phi}(t) \end{pmatrix} = \begin{pmatrix} H_{nw}(k,t) & H_c \\ H_c^\dagger & H_{sc} \end{pmatrix} \begin{pmatrix} \hat{\Psi}_k(t) \\ \hat{\Phi}(t) \end{pmatrix}, \quad k \in [-\pi, \pi], \tag{7}$$

where $\hat{\Phi}(t) = (\dots \hat{\Phi}_{q_{i-1}}^t(t), \hat{\Phi}_{q_i}^t(t), \hat{\Phi}_{q_{i+1}}^t(t), \dots)^t$ with $q_i \in [-\pi, \pi]$. Here we note that the system-bath coupling term $\hat{H}_c$ does not preserve the momentum in Eq. 1. As will be discussed in the next section, this assumption will simplify the calculation of the self-energy correction [1, 2]. Here in Eq. 7, we have

$$H_{sc} = \begin{pmatrix} \ddots & & & & \\ & H_{sc}(q_{i-1}) & 0 & 0 & \\ & 0 & H_{sc}(q_i) & 0 & \\ & 0 & 0 & H_{sc}(q_{i+1}) & \\ & & & & \ddots \end{pmatrix}, \tag{8}$$

where $H_{sc}(q_i) = \epsilon_{q_i}\sigma_0\tau_z - \Delta\sigma_y\tau_y$. Without loss of generality, we assume that the system-bath coupling $V$ is real. Then, we have

$$H_c = \begin{pmatrix} \dots & V\sigma_0\tau_z & V\sigma_0\tau_z & V\sigma_0\tau_z & \dots \end{pmatrix}. \tag{9}$$

Since $H_{nw}(k,t) = H_{nw}(k, t+T)$ is periodic, where $T = 2\pi/\Omega$, based on the Floquet's theorem, the solution of Eq. 7 can be expressed as

$$\begin{pmatrix} \hat{\Psi}_k(t) \\ \hat{\Phi}(t) \end{pmatrix} = e^{-iE_n t} \begin{pmatrix} \hat{u}_n(k,t) \\ \hat{v}_n(t) \end{pmatrix}, \tag{10}$$

where $\hat{u}_n(k,t) = \hat{u}_n(k, t+T)$ and $\hat{v}_n(t) = \hat{v}_n(t+T)$ are periodic functions (or operators), and $E_n$ is the quasienergy. This is similar to the Bloch's theorem. The continuous time translational symmetry is broken, which implies the energy is not conserved. However, they are not totally broken down and the energy can be determined up to $n\Omega$. As a result, the first Floquet Brillouin zone (FBZ) is introduced in the region $[-\Omega/2, \Omega/2]$. Putting Eq. 10 into Eq. 7, one can obtain,

$$\left[\begin{pmatrix} H_{nw}(k,t) & H_c \\ H_c^\dagger & H_{sc} \end{pmatrix} - i\partial_t\right] \begin{pmatrix} \hat{u}_n(k,t) \\ \hat{v}_n(t) \end{pmatrix} = E_n \begin{pmatrix} \hat{u}_n(k,t) \\ \hat{v}_n(t) \end{pmatrix}, \tag{11}$$

Since we know $\hat{u}_n(k,t) = \hat{u}_n(k, t+T)$ and $\hat{v}_n(t) = \hat{v}_n(t+T)$, one can apply the Fourier transformation (FT) to obtain the following time-independent eigenequation written in the Floquet space,

$$\begin{pmatrix} \underline{H_{nw}(k)} & \underline{H_c} \\ \underline{H_c}^\dagger & \underline{H_{sc}} \end{pmatrix} \begin{pmatrix} \hat{u}_n(k) \\ \hat{v}_n \end{pmatrix} = E_n \begin{pmatrix} \hat{u}_n(k) \\ \hat{v}_n \end{pmatrix}, \tag{12}$$

where

$$\left[\underline{H_{nw}(k)}\right]_{mm'} = \frac{1}{T}\int_0^T dt\, e^{im\Omega t}\left[H_{nw}(k,t) - i\partial_t\right]e^{-im'\Omega t}, \quad \left[\underline{H_{sc}}\right]_{mm'} = \frac{1}{T}\int_0^T dt\, e^{im\Omega t}\left[H_{sc} - i\partial_t\right]e^{-im'\Omega t},$$
$$\left[\underline{H_c}\right]_{mm'} = \frac{1}{T}\int_0^T dt\, e^{im\Omega t}\left[H_c\right]e^{-im'\Omega t}, \tag{13}$$

represent the Floquet Hamiltonians of the nanowire, SC, and their couplings, respectively; and

$$[\hat{u}_n(k)]_{m'} = \frac{1}{T}\int_0^T dt\, e^{im'\Omega t}\hat{u}_n(k,t), \qquad [\hat{v}_n]_{m'} = \frac{1}{T}\int_0^T dt\, e^{im'\Omega t}\hat{v}_n(t) \tag{14}$$

are the components of the Floquet wavefunction. An explicit form of $\underline{H_{nw}(k)}$ is

$$\underline{H_{nw}(k)} = \begin{pmatrix} \ddots & & & \\ & \mathcal{H}_{nw}(k) - \Omega\sigma_0\tau_0 & A\sigma_0\tau_z & 0 \\ & A\sigma_0\tau_z & \mathcal{H}_{nw}(k) & A\sigma_0\tau_z \\ & 0 & A\sigma_0\tau_z & \mathcal{H}_{nw}(k) + \Omega\sigma_0\tau_0 \\ & & & & \ddots \end{pmatrix}, \tag{15}$$

where $\mathcal{H}_{nw}(k) = [(-2t_0\cos k - \mu_0)\sigma_0 + V_z\sigma_z + \lambda\sin k\sigma_y]\tau_z$; and $\underline{H_{sc}}$ is

$$\underline{H_{sc}} = \begin{pmatrix} \ddots & & & \\ & H_{sc} - \Omega I_{sc} & 0 & 0 \\ & 0 & H_{sc} & 0 \\ & 0 & 0 & H_{sc} + \Omega I_{sc} \\ & & & & \ddots \end{pmatrix} \tag{16}$$

where $H_{sc}$ is shown in Eq. 8 and $I_{sc}$ is an identity matrix whose dimension is equal to the dimension of $H_{sc}$; and $\underline{H_c}$ is

$$\underline{H_c} = \begin{pmatrix} \dots & V\sigma_0\tau_z I_{sc} & V\sigma_0\tau_z I_{sc} & V\sigma_0\tau_z I_{sc} & \dots \end{pmatrix}. \tag{17}$$

### B. Retarded Green's function and Dyson equation

The retarded Green's function in our dissipative Floquet system can be defined as [3]

$$G_{nw}^R(k; t, t') = -i\theta(t-t')\langle\{\hat{\Psi}_k(t), \hat{\Psi}_k^\dagger(t')\}\rangle, \tag{18}$$

where $\{\hat{\Psi}_k(t), \hat{\Psi}_k^\dagger(t')\} = \hat{\Psi}_k(t)\hat{\Psi}_k^\dagger(t') + \hat{\Psi}_k^\dagger(t')\hat{\Psi}_k(t)$, and $\hat{\Psi}_k(t)$ has been defined in Eq. 2. The Fourier transformation for $\hat{\Psi}_k(t)$ in a periodical driving system can be expressed as

$$\hat{\Psi}_k(t) = \sum_m \int_{-\Omega/2}^{\Omega/2} \frac{d\omega}{2\pi} e^{-i(\omega + m\Omega)t}\hat{\Psi}_{k,m}(\omega). \tag{19}$$

Then the Fourier transformation for the Green's function $G_{nw}^R(k;t,t')$ will be [4]:

$$G_{nw}^R(k;t,t') = \sum_{mn} \int_{-\Omega/2}^{\Omega/2} \frac{d\omega}{2\pi} e^{-i(\omega+m\Omega)t + i(\omega+n\Omega)t'} \left[G_{nw}^R(k,\omega)\right]_{mn}, \tag{20}$$

and

$$\left[G_{nw}^R(k,\omega)\right]_{mn} = \frac{1}{T}\int_0^T dt_{av} \int_{-\infty}^{\infty} dt_{rel} e^{i(\omega+m\Omega)t - i(\omega+n\Omega)t'} G_{nw}^R(k;t,t'), \tag{21}$$

where $t_{av} = \frac{1}{2}(t+t')$ and $t_{rel} = t - t'$.

Note that the Hamiltonian Eq. 12 is a quadratic from. The SC degrees of freedom can be integrated out, which results in the following Dyson equation in the Floquet space (similar treatments can be found in Ref. [1, 3, 5])

$$\underline{G_{nw}^{R,-1}(k,\omega)} = \underline{\omega} - \underline{H_{nw}(k)} - \underline{\Sigma_{sc}(\omega)}. \tag{22}$$

According to Eq. 15-17, the self-energy correction can be expressed as

$$\underline{\Sigma_{sc}(\omega)} = \sum_{q_i} \begin{pmatrix} \ddots & & & & \\ & V^2 \sigma_0\tau_z \cdot Q_{sc}^R(q_i,\omega+\Omega) \cdot \sigma_0\tau_z & 0 & 0 & \\ & 0 & V^2 \sigma_0\tau_z \cdot Q_{sc}^R(q_i,\omega) \cdot \sigma_0\tau_z & 0 & \\ & 0 & 0 & V^2 \sigma_0\tau_z \cdot Q_{sc}^R(q_i,\omega-\Omega) \cdot \sigma_0\tau_z & \\ & & & & \ddots \end{pmatrix}$$

$$:= \begin{pmatrix} \ddots & & & \\ & \Sigma_{sc}(\omega+\Omega) & 0 & 0 \\ & 0 & \Sigma_{sc}(\omega) & 0 \\ & 0 & 0 & \Sigma_{sc}(\omega-\Omega) \\ & & & \ddots \end{pmatrix}, \tag{23}$$

where

$$Q_{sc}^R(q_i,\omega) = \frac{1}{(\omega+i\eta)^2 - (\epsilon_{q_i}^2+\Delta^2)} \begin{pmatrix} (\omega+i\eta)+\epsilon_{q_i} & 0 & 0 & \Delta \\ 0 & (\omega+i\eta)+\epsilon_{q_i} & -\Delta & 0 \\ 0 & -\Delta & (\omega+i\eta)-\epsilon_{q_i} & 0 \\ \Delta & 0 & 0 & (\omega+i\eta)-\epsilon_{q_i} \end{pmatrix} \tag{24}$$

with $\eta = 0^+$ is the static (or equilibrium) Green's function of the SC bath. Now our task is to calculate

$$\Sigma_{sc}(\omega) = V^2 \sigma_0\tau_z \cdot \left[\sum_{q_i} Q_{sc}^R(q_i,\omega)\right] \cdot \sigma_0\tau_z := V^2 \sigma_0\tau_z \cdot q_{bath}^R(\omega) \cdot \sigma_0\tau_z. \tag{25}$$

In order to calculate $q_{bath}^R(\omega)$, we assume the density of states is uniform, which results in

$$q_{bath}^R(\omega) = \sum_{q_i} Q_{sc}^R(q_i,\omega) \simeq \rho_F \int d\epsilon_q Q_{sc}^R(q,\omega). \tag{26}$$

Using the residual theorem, we can obtain

$$q_{bath}^R(\omega) = \pi\rho_F \frac{1}{\sqrt{-(\omega+i\eta)^2+\Delta^2}} \begin{pmatrix} -(\omega+i\eta) & 0 & 0 & -\Delta \\ 0 & -(\omega+i\eta) & \Delta & 0 \\ 0 & \Delta & -(\omega+i\eta) & 0 \\ -\Delta & 0 & 0 & -(\omega+i\eta) \end{pmatrix}$$

$$= \pi\rho_F \frac{1}{\sqrt{-(\omega+i\eta)^2+\Delta^2}} [-(\omega+i\eta) + \Delta\sigma_y\tau_y]. \tag{27}$$

We note that in the literature, the Green's function $q_{bath}^R(\omega)$ is called retarded quasi-classical Green's function [1, 6, 7]. Since the value of $\rho_F$ doesn't matter in our problem, we can simply let $\pi\rho_F = 1$. Therefore,

$$\Sigma_{sc}(\omega) = V^2 \frac{1}{\sqrt{-(\omega+i\eta)^2+\Delta^2}} [-(\omega+i\eta) - \Delta\sigma_y\tau_y]. \tag{28}$$

FIG. 1. Lattice model for the recursive Green function method. We suppose the nanowire is composed of 200 sites, and couple an identical SC bath to each site.

Let $\mathcal{H}_{eff}(k,\omega) = H_{nw}(k) + \Sigma_{sc}(\omega)$, we then get the expression of $G^R_{nw}(k,\omega)$:

$$G^R_{nw}(k,\omega) = \left[\omega - \mathcal{H}_{eff}(k,\omega)\right]^{-1}, \tag{29}$$

where $\mathcal{H}_{eff}(k,\omega)$ is equal to the Eq. 3 in the main text. Our numerical calculation is based on Eq.(29). We truncate $\mathcal{H}_{eff}(k,\omega)$ and just leave a $(21 \times 4) \times (21 \times 4)$ matrix ($N_F = 21$), then numerically calculate $G^R_{nw}(k,\omega)$. Finally, the calculating of $\nu_k(\omega)$ through Eq. (6) in the main text of this paper leads us to Fig. 2 (a) in the main text.

## II. RECURSIVE GREEN'S FUNCTION METHOD

This section will show how to get Fig. 2 (b) in the main body through the recursive Green's function method. For details of the recursive Green's function method, we refer to Ref. [8–10]. In our problem, Green's functions are more structured due to the introducing of Floquet and Keldysh space.

The key to using the recursive Green's function method is to regard the nanowire as a lattice model (see Fig.1). We suppose that this nanowire has 200 sites. Since we just care about the energy spectrum and use the SC bath to produce an induced gap for the nanowire, the correlation in the bath can be ignored under Markovian approximation. In this case, and the influence on each site from the SC bath are the same, which equals to that we couple an identical SC bath to each site.

In order to detect MZMs localized in two edges of this wire, we use the following recursive equation:

$$G^d_{n+1,n+1}(\omega) = \left[G^0_{n+1,n+1}(\omega) - V_{n+1,n} \cdot G^d_{n,n}(\omega) \cdot V_{n,n+1}\right]^{-1} \tag{30}$$

where $V_{n,n+1}$ denotes the hopping term between $n$th and $(n+1)$th site in the Floquet$\otimes$Keldysh$\otimes$BdG space, the subscript $n$ denotes the $n$th site, and the superscript $d$ and $0$ denote the dressed Green's function of one site after considering the hopping term $V_{n,n+1}$ and the bare Green's function of one site with only the onsite energy, respectively. Note that $G^d_{1,1}(\omega) = G^0_{1,1}(\omega)$. After 199 recursions, one will get $G^d_{200,200}(\omega)$. Then using Eq. (9) in the main text of this paper will lead us to Fig. 2 (b) in the main text of this paper.

## III. TOPOLOGICAL PHASE TRANSITION OF FLOQUET MAJORANA ZERO MODES

### A. Comparison with the realistic model

In the main text, we have mentioned that the description of Floquet Majorana zero modes (FMZMs) is also beyond the Floquet topological band theory in the intrinsic SC limit, e.g. Eq. 6 in the main text. Here we use an concrete example to show this. Under the intrinsic SC limit, the Floquet Hamiltonian becomes

$$\mathcal{H}_{int}(k) = \begin{pmatrix} \ddots & & & \\ & \mathcal{H}_{nw}(k) - \Omega - \Delta_{ind}\sigma_y\tau_y & A\sigma_0\tau_z & 0 \\ & A\sigma_0\tau_z & \mathcal{H}_{nw}(k) - \Delta_{ind}\sigma_y\tau_y & A\sigma_0\tau_z \\ & 0 & A\sigma_0\tau_z & \mathcal{H}_{nw}(k) + \Omega - \Delta_{ind}\sigma_y\tau_y \\ & & & & \ddots \end{pmatrix}. \tag{31}$$

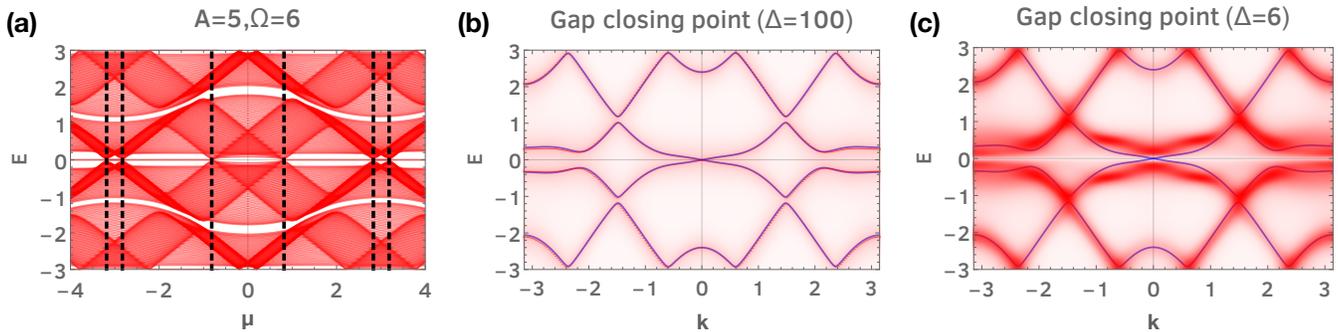

FIG. 2. Topological phase transitions of Floquet Majorana zero modes. (a) shows the result of the intrinsic SC limit, in which the black dashed lines represent the phase transition points. (b)-(c) show the numerical calculation of $\nu_k(\omega)$ without any approximations. As shown in (b), when $\Delta = 100$, which is lager than $(2N_F + 1)\Omega = 66$, $\nu_k(\omega)$ can be approximated described by the Floquet topological band theory in the intrinsic SC limit. Here the parameters are chose as $t = 1, \alpha = 1.5, V_z = 1.2, A = 5, \Omega = 6, N_F = 5$ for (a)-(c), and $\Delta_{ind} = 0.64$ for (a), and $V = 0.8, \mu_c = -3.185, \eta = 0.01$ for (b)-(c).

Using the above Floquet Hamiltonian, one can calculate the open boundary spectrum evolution under the change of external parameters. Choosing $t = 1, \alpha = 1.5, V_z = 1.2, A = 5, \Omega = 6, N_F = 5, \Delta_{ind} = 0.64$, one can obtain the results shown in Fig. 2 (a), where there exist six phase transition points (black dashed lines) in the region $\mu \in [-4, 4]$. Selecting one of the phase transition points, i.e. $\mu_c = -3.185$, one can calculate the periodic boundary Floquet band structure in the first FBZ shown in Fig. 2 (b) with thin blue lines. One can notice that there is a gap closing point at $k = 0$.

Back the original Hamiltonian Eq. 3 in the main text, we can also calculate $\nu_k(\omega)$ with periodic boundary condition. As discussed in the main text, when $\Delta = 100$, which is lager than $(2N_F + 1)\Omega = 66$, $\nu_k(\omega)$ can be approximated by the Floquet band of Eq. 31. Fig. 2 (b) shows their comparisons. In this case, the topological phase transition points can be well described by the Floquet topological band theory. However, as shown in Fig. 2 (c), when $\Delta = 6$, the gap at $k = 0$ opens. This means, the Floquet topological band theory cannot predict the correct topological phase transition points of the FMZM.

## IV. SOME ADDITIONAL NOTES OF THE MAJORANA POISONING MODEL

In this section, we provide some additional calculations of the Majorana poisoning model, whose effective Hamiltonian is,

$$\underline{H(\omega)} = \begin{pmatrix} \ddots & & & \\ & \Sigma(\omega+\Omega) - \Omega & A & 0 \\ & A & \Sigma(\omega) & A \\ & 0 & A & \Sigma(\omega-\Omega) + \Omega \\ & & & & \ddots \end{pmatrix}, \qquad (32)$$

where

$$\Sigma(\omega) = -V^2(\omega + i\eta)/\sqrt{-(\omega+i\eta)^2 + \Delta^2}. \qquad (33)$$

Here $A$, $V$, $\Delta$ represent the driving amplitude, the nanowire-bath coupling strength, and the driving frequency, respectively. The major feature of this effective model is that the Hamiltonian is energy dependent, due to the existence of self-energy correction, $\Sigma(\omega)$. Since for a given $\omega$, $H(\omega)$ is non-Hermitian, it is convenient to introduce the biorthogonal basis [11],

$$\underline{H(\omega)}|u_m^R(\omega)\rangle = E_m(\omega)|u_m^R(\omega)\rangle, \quad \langle u_n^L(\omega)|\underline{H^\dagger(\omega)} = E_n(\omega)\langle u_n^L(\omega)|,$$

where $\langle u_n^L(\omega)|u_m^R(\omega)\rangle = \delta_{nm}$. Based on this basis, one can express the matrix element of the retarded Green's function in the Floquet space,

$$\left[G^R(\omega)\right]_{mn} = \langle m|\frac{1}{\underline{\omega} - \underline{H(\omega)}}|n\rangle = \sum_s \frac{\langle m|u_s^R(\omega)\rangle\langle u_s^L(\omega)|n\rangle}{\omega - E_s(\omega)}, \qquad (34)$$

where $m, n$ represent the Floquet sites, and $s$ represents the eigenvalue index. The time-averaged DoS is defined as the 00-entry of the retarded Green's function defined in the Floquet space, namely,

$$\nu(\omega) = -\frac{1}{\pi} \text{Im} \left[ G^R(\omega) \right]_{00}. \tag{35}$$

Eq. 34 and 35 are the main results of this section. In the following contents, we will analyze their behaviors.

### A. The definition of lifetime

#### 1. Review the definition of lifetime based on the Green's function

We first review the definition of the lifetime based on the Green's function method. Consider a free electron with energy dispersion $E(k)$, the corresponding retarded Green's function can be defined as

$$G_0^R(k, \omega) = \frac{1}{\omega - E(k) + i\eta} \tag{36}$$

Therefore, the corresponding propagator becomes

$$G_0(k, t) = \int_{-\infty}^{\infty} \frac{d\omega}{2\pi} e^{-i\omega t} G_0^R(k, \omega) \propto e^{-iE(k)t - \eta t}. \tag{37}$$

When $\eta > 0$, the propagator of the electron will decay with time, and defines a corresponding lifetime. Note that in the non-interacting isolated systems, $\eta = 0^+$, which means the quasiparticle lifetime is infinity.

Generalizing this argument to the interacting or open quantum systems, the retarded Green's becomes

$$G^R(k, \omega) = \frac{1}{\omega - E(k) - \Sigma(k, \omega)}. \tag{38}$$

The pole of $G^R(k, \omega)$, that is the solution of the following equation

$$\omega - E(k) - \Sigma(k, \omega) = 0, \tag{39}$$

corresponds to the quasiparticle excitation, whose imaginary part is proportional to the inverse of the lifetime. If one only concerns the low energy physics around $\omega = 0$, one can also apply the zero frequency approximation of the self-energy correction, namely, $\Sigma(k, \omega) \simeq \Sigma_0(k) := \Sigma(k, \omega = 0) = \text{Re}[\Sigma_0(k)] - i \text{Im}[\Sigma_0(k)]$ in the calculation. As a result, $E(k) + \text{Re}[\Sigma_0(k)]$ corresponds to the renormalized band structure, and the lifetime of the quasiparticle with momentum $k$ becomes

$$\tau_k = \frac{1}{\text{Im}[\Sigma_0(k)]}. \tag{40}$$

This statement can be understood from the propagator

$$G(k, t) = \int_{-\infty}^{\infty} \frac{d\omega}{2\pi} e^{-i\omega t} G_0^R(k, \omega) \propto e^{-i(E(k) + \text{Re}[\Sigma_0(k)])t - \text{Im}[\Sigma_0(k)]t}, \tag{41}$$

which will decay with time. The character time of this decay is the inversion of $\text{Im}[\Sigma_0(k)]$. On the other hand, the contribution of this mode to the DoS is

$$\nu_k(\omega) = -\frac{1}{\pi} \text{Im} \left[ G^R(k, \omega) \right] = \frac{1}{\pi} \frac{\text{Im}[\Sigma_0(k)]}{(\omega - E(k) - \text{Re}[\Sigma_0(k)])^2 + \text{Im}[\Sigma_0(k)]^2}, \tag{42}$$

which corresponds to a Lorentz peak at $E(k) + \text{Re}[\Sigma_0(k)]$.

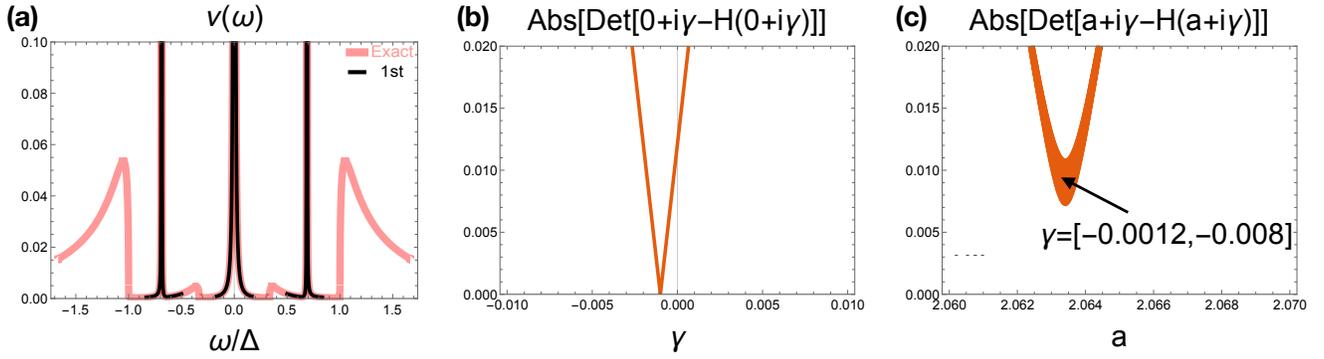

FIG. 3. The peaks and poles in the Floquet Majorana poisoning model. (a) shows the numerical calculation of $\nu(\omega)$ in Eq. 43 with $\Omega = 2$, $\Delta = 3$, $A = 1/2$, $\eta = 1/1000$ and $V = 1$. (b) and (c) shows the function $\text{Abs}[\det[\underline{\omega} - \underline{H}(\omega)]]$ around the peaks at $\omega = 0$ and $\omega \simeq 2.064$.

### 2. Issues of conventional definition of lifetime

In this section, we will show that the conventional definition discussed above cannot simply obtain the lifetime of our Majorana poisoning model. As shown in Fig. 3 (a), $\nu(\omega)$ is plotted with the following parameters $\Omega = 2$, $\Delta = 3$, $A = 1/2$, $\eta = 1/1000$, $N_F = 1$. One can notice that there are three quasiparticle peaks around $\omega = 0, \pm\Omega$ ($\Omega = 2$). In order to understand these three peaks, we first calculate the poles of the retarded Green's function of Eq. 32, i.e.,

$$\det\left[\underline{\omega} - \underline{H}(\omega)\right] = \det\left[\begin{pmatrix} (\omega + \Omega) - \Sigma(\omega + \Omega) & -A & 0 \\ -A & \omega - \Sigma(\omega) & -A \\ 0 & -A & (\omega - \Omega) - \Sigma(\omega - \Omega) \end{pmatrix}\right] = 0. \quad (43)$$

We numerically solve the above equation and find that there only exists one solution, $\omega = 0 + i\eta$. Here we provide an additional numerical verification of this result. We notice that in Fig. 3 (a), the peaks are around $\omega = 0, \pm\Delta$ ($\Delta = 2$). We may expect Eq. 43 has solutions around $\omega = 0, \pm 2$ in the complex plane. As shown in Fig. 3 (b), we plot the following function

$$\text{Abs}\left[\det[\underline{\omega} - \underline{H}(\omega)]\right], \qquad \omega = 0 + i\gamma. \quad (44)$$

One can find that the above function becomes zero at $\gamma = -\eta$. This means $\omega = 0 + i\eta$ is a pole of the Green's function. However, as shown in Fig. 3 (c), when we plot the function

$$\text{Abs}[\det[\underline{\omega} - \underline{H}(\omega)]], \qquad \text{Re}[\omega] \in [2.06, 2.07], \quad \text{Im}[\omega] \in [-0.0012, -0.008], \quad (45)$$

around $\omega = 2$ in the complex $\omega$ plane, the function $\text{Abs}[\det[\underline{\omega} - \underline{H}(\omega)]]$ does not become zero. We note that due to finite $N_F$ namely, $N_F = 1$, the peaks in the second Floquet BZ are not at $\omega = \pm 2$ exactly, but around $\omega = 2.064$. That's the reason why we chose the plot region in (c). The above discussion shows that the lifetime of the quasiparticles in the Majorana poisoning model can not be calculated from the poles of the retarded Green's function.

Now we focus on the quasiparticle peaks around $\omega = 0$. We first illustrate the zero frequency approximation discussed in the subsection III.A.1 also breaks down. As shown in Fig. 4, the dashed lines represents the results of zero frequency approximation, that is

$$\nu(\omega) \simeq \nu_0(\omega) := -\frac{1}{\pi}\text{Im}\left[\frac{1}{\underline{\omega} - \underline{H}(\omega = 0)}\right]_{00}. \quad (46)$$

One can notice that the approximation result $\nu_0(\omega)$ is distinct from the exact result.

### 3. Fitting the quasiparticle peaks

In order to simplify the discussion, we first define the following function

$$\underline{F}(\omega) = \underline{\omega} - \underline{H}(\omega). \quad (47)$$

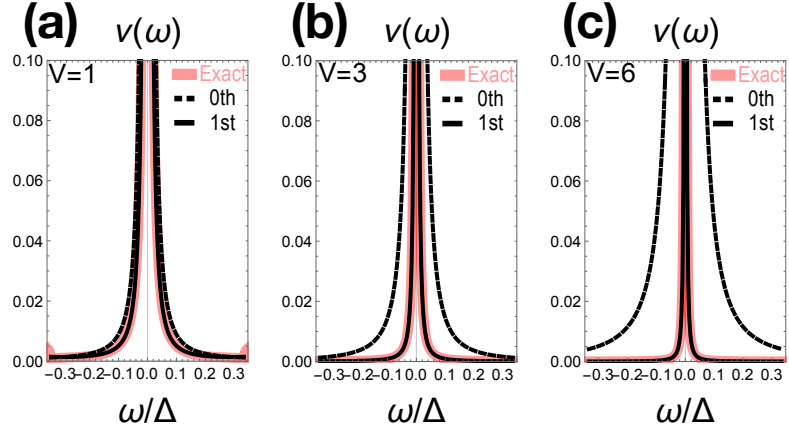

FIG. 4. Floquet Majorana poisoning model. (a)-(c) show the exact, zeroth and first order numerical calculation of $\nu(\omega)$ in Eq. 35 with $\Omega = 2$, $\Delta = 3$, $A = 1/2$, $\eta = 1/1000$, $N_F = 10$, and different values of $V$.

The break-down of the second method discussed above comes from that in $\underline{F}(\omega)$, $\underline{\omega}$ is approximated up to first order, while $\underline{H}(\omega)$ is approximated up to the zeroth order. This inspires us to calculate the following first order expansion of $\omega$ around $\omega = m\Omega$, i.e.,

$$\nu(\omega) \simeq \nu_{m\Omega}^{(1)}(\omega) := -\frac{1}{\pi} \operatorname{Im} \left[ \frac{1}{\underline{F}_{m\Omega}^{(0)} + \underline{F}_{m\Omega}^{(1)}(\delta\omega)} \right]_{00}, \tag{48}$$

where $\delta\omega = \omega - m\Omega$, and the matrix elements of $\underline{F}_{m\Omega}^{(0/1)}$ in the Floquet space with $i,j = -N_F, ..., N_F$ is

$$\left[\underline{F}_{m\Omega}^{(0)}\right]_{ij} = \left[\underline{F}(\omega)\right]_{ij}\Big|_{\omega=m\Omega}, \tag{49}$$

and

$$\left[\underline{F}_{m\Omega}^{(1)}(\delta\omega)\right]_{ij} = \delta\omega \frac{d}{d\omega}\left[\underline{F}(\omega)\right]_{ij}\Big|_{\omega=m\Omega}. \tag{50}$$

Solid black lines in Fig. 4 (a)-(c) represent first order results, which fit well with the exact solution.

### B. Definition of lifetime

Note that peaks in Eq. 48 are determined by the following equation

$$\det\left[\underline{F}_{m\Omega}^{(0)} + \underline{F}_{m\Omega}^{(1)}(\delta\omega)\right] \simeq f_{m\Omega}^{(0)} + \delta\omega f_{m\Omega}^{(1)} + o(\delta\omega^2) = 0, \tag{51}$$

where $f_{m\Omega}^{(\mu)}$ are coefficients of $\delta\omega^\mu$ on the left side of the equation. Therefore, we have $\nu_{m\Omega}^{(1)}(\omega) \propto \operatorname{Im}[1/(f_{m\Omega}^{(0)} + \delta\omega f_{m\Omega}^{(1)})] \propto \operatorname{Im}[1/(f_{m\Omega}^{(0)}/f_{m\Omega}^{(1)} + \delta\omega)] \propto 1/[(\delta\omega - E_{m\Omega})^2 + \Gamma_{m\Omega}^2]$, where $E_{m\Omega} + i\Gamma_{m\Omega} := -f_{m\Omega}^{(0)}/f_{m\Omega}^{(1)}$. Finally, the lifetime of the quasiparticle around $\omega = m\Omega$ becomes

$$\tau|_{\omega=m\Omega} = -\operatorname{Im}[f_{m\Omega}^{(1)}/f_{m\Omega}^{(0)}]. \tag{52}$$

### C. The lifetime of the Floquet Majoranas in different Floquet BZs

Let's first consider a Floquet system without any dissipation. The real wave function can be written as

$$\Psi(t) = \sum_\alpha \lambda_\alpha e^{-i\epsilon_\alpha t}\psi_\alpha(t) \tag{53}$$

where $\psi_\alpha(t) = \psi_\alpha(t+T)$ is the Floquet state for quasi-energy $\epsilon_\alpha$, and their Fourier expansion is $\psi_\alpha(t) = \sum_n e^{-in\Omega t}\psi_\alpha^{(n)}$ with $\Omega = 2\pi/T$. In addition, Floquet theorem tell us the system evolution propagator has a simple form

$$U(t,t') = \sum_\alpha e^{-i\epsilon_\alpha(t-t')}\psi_\alpha(t)\psi_\alpha^*(t'), \tag{54}$$

and therefore, their corresponding Green function [12] can be written as

$$G(n,\omega) = -i\frac{1}{T}\int_0^T dt\, e^{in\Omega t}\int d\tau\, U(t,t-\tau) = \sum_\alpha \sum_k \frac{\psi_\alpha^{(k+n)}(\psi_\alpha^{(k)})^*}{\omega - \epsilon_\alpha - k\Omega + i\delta} \tag{55}$$

where $\delta$ is the infinitesimal regularization. This is the Floquet Green function without considering any dissipation. However, in the presence of dissipation, e.g. due to our self-contained treatment of superconducting proximity effect, we would expect the regularization term becomes a finite dissipation term $\Gamma_k(\omega)$:

$$G(n,\omega) = -i\frac{1}{T}\int_0^T dt\, e^{in\Omega t}\int d\tau\, U(t,t-\tau) = \sum_\alpha \sum_k \frac{\psi_\alpha^{(k+n)}(\psi_\alpha^{(k)})^*}{\omega - \epsilon_\alpha - k\Omega + i\Gamma_k(\omega)}. \tag{56}$$

Our numerical calculation shows that the Floquet Majorana peaks in different Floquet BZs have different height and width (see Fig. 2 in the main text), which leads us to think if those dissipation terms $\Gamma_k(\omega)$ are the same or not for different Floquet BZs (or different $k$).

Now we show that all the quasiparticles in different Floquet BZs have the same dissipation or lifetime. We notice that due to the discrete time translational symmetry, we have

$$\det\left[\underline{\omega} - \underline{H}(\omega)\right] = \det\left[(\omega + m\Omega) - \underline{H}(\omega + m\Omega)\right]. \tag{57}$$

Therefore, the expansion of $\det[\underline{\omega} - \underline{H}(\omega)]$ around $\omega = m\Omega$ must share the same expression. Therefore, their lifetimes are the same. This is consistent with the intuition, since the total Hamiltonian does not break the discrete time translational symmetry. Therefore, we conclude all $\Gamma_k(\omega)$ for different $k$s are the same, but the numerator in the Floquet Green function could be different.

### D. The differences between the peaks in different Floquet BZs

This section explains the differences between the FMZMs in different Floquet bands. We start from the $mn$-entry of the retarded Green's function

$$\left[\underline{G^R(\omega)}\right]_{mn} = \langle m|\frac{1}{\underline{\omega} - \underline{H}(\omega)}|n\rangle = \sum_s \frac{\langle m|u_s^R(\omega)\rangle\langle u_s^L(\omega)|n\rangle}{\omega - E_s(\omega)}. \tag{58}$$

Here $\underline{H}(\omega)|u_s^R(\omega)\rangle = E_s(\omega)|u_s^R(\omega)\rangle$, $\langle u_s^L(\omega)|\underline{H^\dagger(\omega)} = E_s(\omega)\langle u_s^L(\omega)|$, and $\langle u_n^L(\omega)|u_m^R(\omega)\rangle = \delta_{nm}$ is the biorthogonal basis. The time-averaged DoS is defined by

$$\nu(\omega) = -\frac{1}{\pi}\text{Im}\left[\underline{G^R(\omega)}\right]_{00} = -\frac{1}{\pi}\text{Im}\left[\sum_s \frac{\langle 0|u_s^R(\omega)\rangle\langle u_s^L(\omega)|0\rangle}{\omega - E_s(\omega)}\right]. \tag{59}$$

As discussed in the main text, the wavefunctions $\langle 0|u_s^R(\omega)\rangle$ are localized in the central part of the Floquet base. Therefore, the DoS will decay in the higher Floquet BZs.

### E. Comparison with the realistic model

Our model mimics the physical origin of the dissipation of Floquet Majoranas. Once the existence of Floquet Majoranas are identified by the numerical calculations, e.g. Fig. 2 in the main text, we believe that the corresponding lifetime of these Floquet Majoranas can be estimated by our effective model. We first numerically verify the above statement. Here we label the LDOS of the realistic model as $\nu_r$, whose parameters are chosen the same as Fig. 2 (b3) in the main text; label the effective model of FMZMs as $\nu_{e,0}$; and label the effective model of FMPMs as $\nu_{e,\pi}$. As

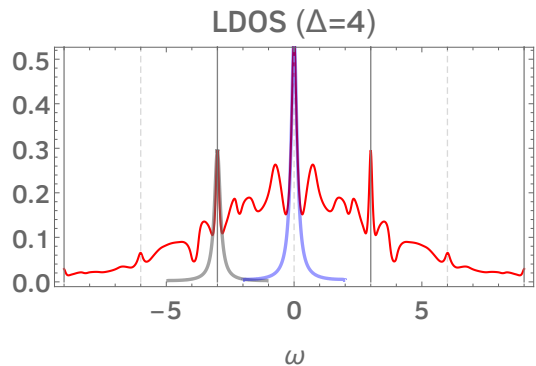

FIG. 5. Comparison of the local density of states (LDOS) of the realistic model (red), Floquet Majorana poisoning model of the zero mode (light blue) in the region $\omega \in [-2, 2]$, and Floquet Majorana poisoning model of the pi mode (light green) in the region $\omega \in [-1, -5]$.

shown in Fig. 5, we numerically found that the peaks of $\nu_r$ (red) are similar to the modified LDOS $\nu_{e,0}/3.27$ around $\omega = 0$ (light blue) and $\nu_{e,\pi}/6.1$ around $\omega = \Omega/2 = 3$ (gray). This means the lifetime of the FMZM can be well approximated by our effective model.

In order to estimate the lifetime in our model, we need to know the parameters achieved in recent experiments, which include the external SC gap $\Delta$, the system-bath coupling strength $V$, the driving amplitude $A$ and frequency $\Omega$, and the $\eta$ appeared in the self-energy $\Sigma(\omega) = -V^2(\omega + i\eta)/\sqrt{-(\omega + i\eta)^2 + \Delta^2}$. In general, the superconductivity gap of Al is about $\Delta = 0.2$meV; the system-bath coupling strength typically depends on the realistic devices, whose value can be hardly identified exactly; and the driving parameters $A$ and $\Omega$ are experimentally tunable. Therefore, we can rescale the parameters based on the external SC gap, which results $\Delta = 0.2$meV, $V = 0.04$meV, $A = 0.075$meV, $\Omega = 0.3$meV, and $\eta = 0.0025$meV (regularization). Under the current parameters, the inverse of the lifetime (or dissipation) of the FMZM is about $0.0069$meV. We emphasize that the dissipation of FMZM could be made much smaller if we consider a weaker semiconductor-SC coupling with also a smaller regularization parameter. Our key point of the paper is to show the lifetime of FMZM in general is finite, even if their equilibrium counterparts may have infinite lifetime. The generalization to the FMPM is straightforward. Using the same parameters, the dissipation of FMPM is the same.

[1] D. E. Liu, A. Levchenko, and R. M. Lutchyn, Phys. Rev. B **95**, 115303 (2017).
[2] T. D. Stanescu and S. Das Sarma, Phys. Rev. B **96**, 014510 (2017).
[3] T. Kitagawa, T. Oka, A. Brataas, L. Fu, and E. Demler, Phys. Rev. B **84**, 235108 (2011).
[4] H. Aoki, N. Tsuji, M. Eckstein, M. Kollar, T. Oka, and P. Werner, Rev. Mod. Phys. **86**, 779 (2014).
[5] *Many-Particle Physics* (1981).
[6] J. Rammer and H. Smith, Rev. Mod. Phys. **58**, 323 (1986).
[7] V. Chandrasekhar, "An introduction to the quasiclassical theory of superconductivity for diffusive proximity-coupled systems," (2003), arXiv:cond-mat/0312507 [cond-mat.mes-hall].
[8] D. J. Thouless and S. Kirkpatrick, Journal of Physics C: Solid State Physics **14**, 235 (1981).
[9] P. A. Lee and D. S. Fisher, Phys. Rev. Lett. **47**, 882 (1981).
[10] P. S. Drouvelis, P. Schmelcher, and P. Bastian, Journal of Computational Physics **215**, 741 (2006).
[11] D. C. Brody, Journal of Physics A: Mathematical and Theoretical **47**, 035305 (2013).
[12] S. Kohler, J. Lehmann, and P. Hänggi, Physics Reports **406**, 379 (2005).



\end{document}